\documentclass[12pt]{article}
\usepackage[latin1]{inputenc}
\usepackage{cite}
\usepackage{amsmath}
\usepackage{amsfonts}
\usepackage{amssymb}
\usepackage{graphicx}
\usepackage{geometry}
\usepackage{amssymb,epsfig}
\usepackage{hyperref}
\usepackage{slashed,caption}
\makeatletter
\renewcommand\section{\@startsection {section}{1}{\z@}%
                                 {-3.5ex \@plus -1ex \@minus -.2ex}%nn
                                   {2.3ex \@plus.2ex}%
                                   {\normalfont\large\bfseries}}
\renewcommand\subsection{\@startsection{subsection}{2}{\z@}%
                                   {-3.25ex\@plus -1ex \@minus -.2ex}%
                                     {1.5ex \@plus .2ex}%
                                     {\normalfont\bfseries}}
\renewcommand\subsubsection{\@startsection{subsubsection}{3}{\z@}%
                                   {-3.25ex\@plus -1ex \@minus -.2ex}%
                                     {1.5ex \@plus .2ex}%
                                     {\normalfont\itshape}}
\makeatother

\def\pplogo{\vbox{\kern-\headheight\kern -29pt
\halign{##&##\hfil\cr&{\ppnumber}\cr\rule{0pt}{2.5ex}&\ppdate\cr}}}
\makeatletter
\def\ps@firstpage{\ps@empty \def\@oddhead{\hss\pplogo}%
  \let\@evenhead\@oddhead % in case an article starts on a left-hand page
}%      The only change in \maketitle is \thispagestyle{firstpage} instead of \thispagestyle{plain}
\def\maketitle{\par
 \begingroup
 \def\thefootnote{\fnsymbol{footnote}}
 \def\@makefnmark{\hbox{$^{\@thefnmark}$\hss}}
 \if@twocolumn
 \twocolumn[\@maketitle]
 \else \newpage
 \global\@topnum\z@ \@maketitle \fi\thispagestyle{firstpage}\@thanks
 \endgroup
 \setcounter{footnote}{0}
 \let\maketitle\relax
 \let\@maketitle\relax
 \gdef\@thanks{}\gdef\@author{}\gdef\@title{}\let\thanks\relax}
\makeatother

\numberwithin{equation}{section}

\newcommand{\be}{\begin{equation}}
\newcommand{\bea}{\begin{eqnarray}}
\newcommand{\ee}{\end{equation}}
\newcommand{\eea}{\end{eqnarray}}

\newcommand{\tomegap}{\tilde\omega_p}
\newcommand{\tomegaq}{\tilde\omega_q}
\newcommand{\tomegapq}{\tilde\omega_{pq}}

\begin{document}
 
\setcounter{page}0
\def\ppnumber{\vbox{\baselineskip14pt
%\hbox{hep-th/0000000}
}}
\def\ppdate{\footnotesize{}} \date{}

\author{Carlos Tamarit\\
[7mm]
{\normalsize  \it Perimeter Institute for Theoretical Physics, Waterloo, ON, N2L 2Y5, Canada}\\
[3mm]
{\tt \footnotesize  ctamarit at perimeterinstitute.ca }
}

%\bigskip
\title{\bf Higgs vacua with potential barriers}
\maketitle

\begin{abstract} \normalsize
\noindent 
Scenarios in which the Higgs vacuum arises radiatively and is separated from the origin by a potential barrier  at zero temperature are known to be attainable in models with extra singlet scalars, which in the limit of zero barrier height give rise to Coleman-Weinberg realizations of electroweak symmetry breaking.  However, this requires large values of Higgs-portal couplings or a large number $N$ of singlets. This is quantified in detail by considering, for varying $N$, the full two-loop effective potential at zero temperature, as well as finite temperature effects including the dominant two-loop corrections due to the singlets. Despite the large couplings, two-loop effects near the electroweak scale are under control, and actually better behaved in models with larger couplings yet fewer singlets. Strong first-order phase transitions are guaranteed even in the Coleman-Weinberg scenarios. Cubic Higgs  couplings and Higgs associated-production cross sections exhibit deviations from the Standard Model predictions which could be probed at a linear collider.
\end{abstract}

\newpage
 %%%%%%%%%%%%%%%%%%%%%%%%%%%%
 %%%%%%%%%%%%%%%%%%%%%%%%%%%%
 %%%%%%%%%%%%%%%%%%%%%%%%%%%%
 %%%%%%%%%%%%%%%%%%%%%%%%%%%%
 \section{Introduction}
 
 Little is known about the self-interactions and potential of the recently discovered Higgs boson \cite{Aad:2012tfa,Chatrchyan:2012ufa}. In the Standard Model (SM), it is assumed that electroweak symmetry breaking is a tree-level effect triggered by the interplay between a negative mass term and a positive quartic coupling. Although extremely simple, this picture is not free of drawbacks. Foregoing the infamous naturalness problem, it is known that for the observed Higgs mass of 125 GeV the electroweak phase transition in the early Universe fails to be of the strong first-order type \cite{Bochkarev:1987wf}, which rules out the possibility that the SM may explain the baryon asymmetry by the mechanism of electroweak baryogenesis \cite{Kuzmin:1985mm} (see \cite{Morrissey:2012db} for a review). 
 Furthermore, an extrapolation of the Standard Model effective potential to large values of the field reveals that the electroweak vacuum is metastable (see ref.~\cite{Degrassi:2012ry} and references therein), which can be understood from the properties of the running Higgs quartic coupling, which becomes negative at large scales due to the interactions of the top quark. 
 
 These issues are suggestive of new physics correcting the problematic properties of a Standard Model Higgs at 125 GeV, both ensuring a strong first-order phase transition and improving the stability of the Higgs potential. These two things can be accomplished by considering additional scalar fields interacting with the Higgs. Such models have been the subject of intense research, since in their minimality they can address a host of the Standard Model shortcomings, not only stability and baryogenesis but also providing possible dark matter candidates \cite{McDonald:1993ex,Burgess:2000yq,Davoudiasl:2004be,Barger:2007im,Barger:2008jx}.

 Indeed, Higgs-portal interactions involving new scalars can cure the instability by providing positive contributions to the beta function of the Higgs quartic coupling \cite{Gonderinger:2009jp}, or by means of threshold effects \cite{EliasMiro:2012ay}. Additionally, they can make the electroweak phase transition of the strong first-order type  by enhancing the potential barrier between the origin and the electroweak vacuum, either by purely thermal or zero-temperature effects. An enhancement by thermal effects is the mechanism usually considered in the literature, relying on the  temperature-dependent cubic Higgs interactions that  bosonic fields --typically new scalars in singlet or supersymmetric extensions of the SM-- contribute to the thermal potential \cite{Anderson:1991zb,Espinosa:1993bs,Profumo:2007wc,Carena:1996wj,Cohen:2011ap}. Another possibility entails the generation of a barrier in the effective potential already at zero temperature. Under the assumption that all fields other than the Higgs have zero vacuum expectation values (VEVs), in these scenarios the sign of the tree-level parameters in the Higgs potential are reversed with respect to the usual SM picture: the Higgs potential is concave at the origin, implying a positive Higgs quadratic term, and a barrier is generated by a negative quartic.  With a negative quartic the potential would be unbounded from below, so that new physics has to be invoked to generate the electroweak vacuum and stabilize the theory. In order to keep the barrier, the stabilization cannot happen at tree-level, implying that the electroweak vacuum has to arise from quantum effects. This immediately suggests a connection with Coleman-Weinberg realizations of electroweak symmetry breaking, in which the Higgs vacuum is triggered dynamically by quantum effects in the absence of tree-level mass parameters, avoiding the naturalness problem by providing a dynamical origin for the mass-scales in the theory \cite{Coleman:1973jx,Gildener:1976ih}. Indeed, Coleman-Weinberg scenarios can be obtained as a limiting case of the models with zero-temperature barriers when the barrier height (related to the Higgs mass parameter) goes to zero, if such models allow for turning off all the other mass parameters.
 
 Scenarios with a zero-temperature barrier in the Higgs potential were introduced in ref.~\cite{Grojean:2004xa}, with new physics modeled by a higher dimensional $h^6$ operator; see refs.~\cite{Ham:2004zs,Bodeker:2004ws,Delaunay:2007wb,Grinstein:2008qi} for related works. Given that the new physics scale associated with the higher dimensional operator cannot be far from the weak scale (as is clear from the fact that this operator has to generate the electroweak vacuum), it becomes relevant to study ultraviolet completions. Indeed, they do exist in models with additional singlet scalars, introduced in ref.~\cite{Espinosa:2007qk} and further analyzed in ref.~\cite{Espinosa:2008kw}.  These works also stressed the connection with Coleman-Weinberg realizations of electroweak symmetry breaking alluded to before; for related works in classically scale-invariant models in which fields other than the Higgs have suppressed VEVs, see also~\cite{Foot:2007as,AlexanderNunneley:2010nw,Dermisek:2013pta,Guo:2014bha} (for other possible Coleman-Weinberg constructions, involving nonzero VEVs for other fields other than the Higgs, see for example refs.~\cite{Meissner:2006zh,Foot:2007iy,Hempfling:1996ht,Chang:2007ki,Iso:2009ss,Iso:2012jn,Englert:2013gz,Hambye:2013dgv,Carone:2013wla,Khoze:2014xha,Abel:2013mya}). 
 
 The scenarios of refs.~\cite{Espinosa:2007qk,Espinosa:2008kw} have a large number of real scalars (12) with sizable $\lambda_{SH}\sim O(1)$ Higgs-portal couplings. This opens up the question of whether loop corrections involving the Higgs-portal coupling, which are expected to go as $N\lambda_{SH}$ ($N$ being the number of singlets), are under control, as well as whether models with fewer scalars are viable, and how the required Higgs-portal couplings change with the number of scalars. To achieve equivalent effects in the Higgs potential with fewer singlets, one expects even larger values of $\lambda_{SH}$, which may further challenge perturbative stability.
 
 In order to answer these questions, as well as revisit some of the phenomenological implications in view of the now known value of the Higgs mass, this paper centers on the study of scenarios \`{a} la ref.~\cite{Espinosa:2007qk}, involving  additional scalars transforming as singlets under the SM gauge group, in which the Higgs vacuum is generated radiatively together with a potential barrier at zero temperature, ensuring a strong first-order phase transition and providing realizations of Coleman-Weinberg electroweak symmetry breaking in the limit of zero barrier height. The singlets, taken as complex, are assumed to be stabilized with a zero or small VEV, suppressing the mixing with the Higgs. Additionally, if the singlet sector has a global symmetry, zero singlet VEVs prevent light Goldstone modes in the spectrum, making it unnecessary to have the singlets charged under a hidden gauge group.\footnote{For models allowing for nonzero VEVs for the new fields and having tree-level vacua with zero-temperature barriers, see for example ref.~\cite{Espinosa:2011ax}.  Also, allowing for additional VEVs and  hidden-sector gauge fields one may avoid the need for strong coupling \cite{Hambye:2013dgv}. I thank A. Strumia for correspondence on this point.} In this work, the properties of these models are studied in detail for different values of the number $N$ of complex singlets and the Higgs-portal coupling $\lambda_{SH}$, extending the results of refs.~\cite{Espinosa:2007qk,Espinosa:2008kw} for arbitrary  $N$ and clearly establishing the need for  large $\lambda_{SH}$ or large singlet multiplicity. Given the need of large couplings, the analysis in refs.~\cite{Espinosa:2007qk,Espinosa:2008kw} is improved by considering the full renormalization-group improved effective potential at two-loops, as well as including the dominant two-loop effects involving the Higgs-portal coupling in the finite temperature effective potential. 
 
 The results show that the value of the Higgs-portal coupling $\lambda_{SH}$ needed to achieve a certain barrier height goes as $1/\sqrt{N}$, with $\lambda_{SH}\gtrsim6.5$ for $N=1$. Remarkably, two-loop corrections are well-behaved even for $N=1$, as evidenced by studying the behavior of the renormalization-group-improved effective potential under changes of scale. This behavior is actually better for small $N$, which may be taken as evidence that the dominant perturbative corrections essentially depend on $\lambda_{SH} N\sim\sqrt{N}$. Hence,  models with a small number of singlets are not only viable but also better behaved. The finite-temperature computations confirm that strong first-order phase transitions are attainable for all $N$ even in the limiting Coleman-Weinberg scenarios. It is also shown that, although direct signals of singlet production (such as jets plus missing energy) are well below current limits, deviations from the Standard Model in effective cubic Higgs interactions and Higgs associated production cross-sections are sizable and potentially measurable at linear colliders. Cubic Higgs interactions are typically enhanced by a factor of two, in line with the expectations of ref.~\cite{Noble:2007kk} for generic models with a strong first-order electroweak phase transition, and  larger than previous estimates ~\cite{Espinosa:2008kw}. On the other hand, deviations in Higgs associated production cross sections are estimated to be of the order of 1.4\% and greater.
 
 The organization of this paper is as follows. First, the construction of the two-loop, renormalization-group improved effective potential is reviewed in \S~\ref{sec:VCW}, in which the behavior of perturbation theory is examined and results tying the height of the barrier with the value of the Higgs-portal coupling are presented. \S~\ref{sec:FT} deals with finite temperature effects and the strength of the electroweak phase transition, while collider and cosmological constraints are  examined in \S~\ref{sec:pheno}. Two appendices are included, \S~\ref{app:RG} providing two-loop beta functions for the SM supplemented with a complex multiplet with a $U(N)$ global symmetry, and \S~\ref{app:diagrams} listing formulae for the temperature-dependent part of two-loop scalar diagrams involving the Higgs-portal coupling.
 
 %%%%%%%%%%%%%%%%%%%%%%%%%%%%
 %%%%%%%%%%%%%%%%%%%%%%%%%%%%
 %%%%%%%%%%%%%%%%%%%%%%%%%%%%
 %%%%%%%%%%%%%%%%%%%%%%%%%%%%
 \section{Two-loop renormalization-group improved effective potential in singlet extensions of the Standard Model\label{sec:VCW}}
 
 Let's consider the Standard Model supplemented with a complex, $N$-dimensional scalar multiplet $S$ behaving as a singlet under the Standard Model gauge group. Assuming for simplicity that $S$ is charged under a $U(N)$ global symmetry, the most general renormalizable potential compatible with the symmetries is of the form
 \begin{align}
 \label{eq:Vscalar}
  V^{\rm tree}=&m^2_H H^\dagger H+m^2_S S^\dagger S+\frac{\lambda}{2}(H^\dagger H)^2+\frac{\lambda_S}{2}(S^\dagger S)^2+\lambda_{SH} (H^\dagger H)(S^\dagger S).
  %%%%%
 \end{align}
 Although interactions that violate the $U(N)$ symmetry are interesting in their own right, since for example they might source new CP-violating phases which may help generate the baryon asymmetry, they will be ignored here for simplicity under the assumption that their effects are subdominant when it comes to radiatively generate a Higgs vacuum with a potential barrier.  This assumption is partly motivated by the desire to have $S$ stabilized at the origin, which would not be possible in the presence of large couplings that violate the $U(N)$ symmetry and introduce runaway directions for the singlets, as happens for example with the interactions $\kappa_{ij}(H^\dagger H) (S_iS_j+c.c.)$. Thus one may set $S=0$ in the effective potential and only consider its dependence on the Higgs field, as will be done in what follows.
 If one denotes the set of couplings of the theory as $\{\delta_i\}$ and the renormalization scale as $\mu$, the effective potential  for the real part of the neutral component of the Higgs field, $h=\sqrt{2}\,\text{Re}H^0$, will be of the form (see ~\cite{Quiros:1999jp} for a review):
 \begin{align}
 \label{eq:V}
 V(h,\delta_i,\mu)\equiv&\Omega(\delta_i,\mu)+V_h(h,\delta_i,\mu)=\\
  %%%
  \nonumber=&V^{\rm tree}(h,\delta_i,\mu)+\frac{1}{16 \pi^2}V^{(1)}(h,\delta_i,\mu)+\frac{1}{(16 \pi^2)^2}V^{(2)}(h,\delta_i,\mu).
 \end{align}
 In the above equation, $\Omega$ represents the vacuum energy, while $V_h$ includes the nontrivial field dependence. The one and two-loop contributions $V^{(1)}$ and $V^{(2)}$  can be calculated in terms of the mass matrices and couplings of the theory expressed in terms of real scalars and Weyl fermions in the background of the $h$ field. The one-loop contribution is given by 
 \begin{align}\nonumber
V^{(1)}(h,\delta_i,\mu) =&\sum_{a}\frac{3}{4}(m^2_a(h))^2\left[\log\frac{m^2_a(h)}{\mu^2}-\frac{5}{6}\right]+\sum_{i}\frac{1}{4}(m^2_i(h))^2\left[\log\frac{m^2_i(h)}{\mu^2}-\frac{3}{2}\right]\\
%%%%
&-\sum_{I}\frac{1}{2}(m^2_I(h))^2\left[\log\frac{m^2_I(h)}{\mu^2}-\frac{3}{2}\right],\label{eq:V1loop}
 \end{align}
 where the indices $a,i,I$ label vector fields, real scalars and Weyl fermions, and $m^2_k(h)$ denote mass eigenvalues in the background of the field $h$. Writing the Higgs field as $H=(\frac{1}{\sqrt{2}}(h^+_r+ih^+_i),\frac{1}{\sqrt{2}}(h+i\chi))^\intercal$ and denoting the Goldstone fields as $G_i=\chi, h^+_r,h^+_i$, then the tree-level field-dependent masses are given by (using the Grand Unified Theory (GUT) renormalization of the hypercharge coupling):
 \begin{align}
  m^2_W(h)=&\frac{  g_2^2h^2}{4},
& m^2_Z(h)=& \left(\frac{3 g_1^2}{5}+g_2^2\right)\frac{ h^2}{4}, &
 \label{eq:massesh} m^2_h(h)=&m^2_h+\frac{3\lambda h^2}{2},\\
 %%%%
  \nonumber  m^2_G(h)=&m^2_h+\frac{\lambda h^2}{2}, & m^2_S(h)=&m^2_S+\frac{\lambda_{SH} h^2}{2}.
 \end{align}
The reader is referred to ref.~\cite{Martin:2001vx} for details on the two-loop contributions.

The perturbative expansion of the effective potential in eq.~\eqref{eq:V} can be improved by using RG techniques, giving rise to the renormalization-group improved effective potential, which allows to resum logarithmic corrections as well as motivate choices of the renormalization scale leading to accurate calculations for all values of $h$. The field-dependent contribution $V_h$ to the effective potential in eq.~\eqref{eq:V} satisfies the Callan-Symanzik equation
\begin{align}
\label{eq:CallanSymanzik}
 \left[\mu\frac{\partial}{\partial\mu}+\sum \beta_{\delta_i}\frac{\partial}{\partial\delta_i}-\gamma_Hh\frac{\delta}{\delta h}\right]V_h=0 ,
\end{align}
where $\beta_{\delta_i}=\frac{\mu\partial\delta_i}{\partial_\mu}$ and $\gamma_H=\frac{1}{2 Z_H}\frac{\mu\partial Z_H}{\partial_\mu}$ are the beta functions and Higgs anomalous dimension of the theory, whose two-loop values are given in appendix~\ref{app:RG}. Eq.~\eqref{eq:CallanSymanzik} implies that $V_h$ can be written as
\begin{align}
\label{eq:RGimproved}
 V_h(h,\delta_i,\mu)=V_h(\xi(t)h,\hat\delta_i(t),\hat\mu(t)),
\end{align}
where $t$ is an arbitrary parameter associated with the freedom to perform rescalings without altering the physics, such that $\hat\mu(t)=\mu e^t$, while $\xi(t)$ and $\hat\delta_i(t)$ are the running field renormalization factor and couplings, satisfying the Renormalization Group (RG) equations
 \begin{align*}
  \frac{1}{\xi(t)}\frac{d\xi(t)}{dt}=-\gamma_H(\delta_i(t)),\quad \xi(0)=1,\\
  %%%
    \frac{d\hat\delta_j(t)}{dt}=\beta_{\delta_j}(\hat\delta_i(t)), \quad \hat\delta_i(0)=\delta_i.
 \end{align*}
 Indeed, the Callan-Symanzik equation \eqref{eq:CallanSymanzik} is equivalent to 
\begin{align*}
 \frac{d}{dt}V_h(\xi(t)h,\hat\delta_i(t),\hat\mu(t))=0,
\end{align*}
implying invariance under rescalings. Note that this refers to $V_h$ rather than the full potential including the vacuum energy, $V=\Omega+V_h$. In general, one will have 
\begin{align*}
 \left[\mu\frac{\partial}{\partial\mu}+\sum \beta_{\delta_i}\frac{\partial}{\partial\delta_i}-\gamma_Hh\frac{\delta}{\delta h}\right]V=\left[\mu\frac{\partial}{\partial\mu}+\sum \beta_{\delta_i}\frac{\partial}{\partial\delta_i}\right]\Omega,
\end{align*}
implying that the full potential (in particular the vacuum energy) is not scale-invariant. This is not a problem unless one were to perform field-dependent changes of scale in $V$, i.e. $t=t(h)$, since in this case $\Omega$ and $V_h$ will become mixed and the shape of the potential won't be invariant under the rescaling. Now, field-dependent choices of scale are motivated by the desire to get accurate results for large values of $h$. The loop corrections to the effective potential involve logarithms of the type $\log\frac{m^2_j(h)}{\mu^2}$, where $m^2(h_j)$ denotes an eigenvalue of any of the mass matrices of the fields in the background $h$. Thus, it is clear that for large values of $h$, in which $m^2_j(h)\sim h^2$, perturbative corrections are minimized by choosing $\mu\sim h$. If this field-dependent rescaling is allowed, then the large-field behavior of the effective potential will be captured by  the running couplings evaluated at the scale of the field.\footnote{This reasoning lies behind the well-known results tying the instability of the SM potential at large values of the Higgs field with negative values of the running quartic coupling.} To allow such a field-dependent rescaling without changing the shape of the potential, one may redefine the vacuum by subtracting  a contribution $\Delta\Omega$ such that
\begin{align*}
\left[\mu\frac{\partial}{\partial\mu}+\sum \beta_{\delta_i}\frac{\partial}{\partial\delta_i}\right](\Omega+\Delta\Omega)=0.
\end{align*}
In this case the full potential will satisfy an equation analogous to \eqref{eq:RGimproved}, allowing field-dependent rescalings  without altering the physics, so that by choosing $t=f(h)$ the resulting effective potential will be accurate for all values of $h$. If the potential $V$ is computed at a loop order $L$, then substituting $h$ for $\xi(t)h$ as in eq.~\eqref{eq:RGimproved} and calculating the couplings $\delta_i(t)$ with $L+1$ accuracy is known to resum all leading logarithmic corrections of order $L+1$; for this reason the resulting potential is known as the RG-improved effective potential. 

For the theory at hand, a computation of the two-loop effective potential from the formulae in ref.~\cite{Martin:2001vx} yields, in the limit in which the gauge couplings are neglected,
\begin{align*}
 \Omega(\mu,\delta_i)=&\frac{1}{16\pi^2} \left[m^4{}_H \left(\log \left(\frac{m^2{}_H}{\mu ^2}\right)-\frac{3}{2}\right)+N m^4{}_S \left(\frac{1}{2} \log \left(\frac{m^2_S}{\mu ^2}\right)-\frac{3}{4}\right)\right]\\
 &+\frac{1}{(16\pi^2)^2}\left[2 N m^2{}_H m^2_S \lambda _{SH} \left(\log \left(\frac{m^2{}_H}{\mu ^2}\right)-1\right) \left(\log \left(\frac{m^2_S}{\mu ^2}\right)-1\right)\right.\\
 %%%N
 &\left.+3 \lambda  m^4{}_H \left(\log \left(\frac{m^2{}_H}{\mu ^2}\right)-1\right){}^2+ \frac{1}{2}N(N+1)m^4{}_S \lambda _S \left(\log \left(\frac{m^2_S}{\mu ^2}\right)-1\right){}^2\right].
\end{align*}
For a given choice of the couplings $m^2_S,\lambda_S, \lambda_{SH}$, one may fix $m^2_H$ and $\lambda$ from the measurements of the SM gauge couplings and the known values of the masses of the fermions, gauge bosons and Higgs. The presence of extra scalars implies in principle some modifications in the usual matching relations between experimental observations and the theory's couplings. However, there are no  modifications from the Standard Model matching relations at one-loop order \footnote{This is related with the fact that the one-loop beta functions of gauge and Yukawa couplings are identical to those in the Standard Model, see \S~\ref{app:RG}.}, so that the SM results will be used.
The values of the Standard Model gauge couplings are taken from the Particle Data Group tables \cite{Beringer:1900zz}, which are used to calculate the SM gauge couplings at the scale of the mass of the Z boson in the $\overline{\rm MS}$ scheme. These are evolved  up to the scale of the top mass at 2-loop precision with the 2-loop SM RG equations \cite{Luo:2002ey}, starting with tree-level estimates for the initial values of the top and bottom Yukawas and a guessed value for the Higgs quartic coupling. At $\mu=m_t$, the values of  $y_t$ and $y_b$ are recalculated from the pole masses of the quarks by applying QCD threshold corrections of order $\alpha_s^2$ \cite{Chetyrkin:1999qi}. Things are evolved again up to the scale of the Higgs vacuum expectation value $v(\mu=m_Z)=246\, {\rm GeV}$, at which $m^2_S,\lambda_S, \lambda_{SH}$ are given some chosen values and then $\lambda$ and $m^2_H$ are determined by imposing  the following conditions  on the effective potential:
\begin{align}\label{eq:lambdam2H}
\left.\begin{array}{c}
 \frac{d}{dh}V(\xi(t)h,\hat\delta_i(t),\hat\mu(t))=0 \\
 %%%
  \frac{d^2}{dh^2}V(\xi(t)h,\hat\delta_i(t),\hat\mu(t))=m^2_h+\Delta \Pi(m^2_h)
 \end{array}\right\}
 \text{ for }\mu=v(m_z),t=0,\,\, h=\frac{v(m_Z)}{\xi(\log\frac{m_Z}{v(m_Z)})}\equiv \tilde v.
\end{align}
The first condition guarantees that the potential at the scale $v(m_Z)$ has a minimum at a value of the field equaling  $v(m_Z)$ times the field renormalization factor appropriate from going to the scale $m_Z$ to the scale $v(m_Z)$. The second condition ensures that the mass of the Higgs matches the experimental value $m_h=125$ GeV. The physical mass corresponds to the pole mass, determined by the two-point  function $\Pi(p^2)$ evaluated at $p^2=m^2_h$, while the effective potential only captures the two-point function at zero momentum; this explains the origin of the term $\Delta \Pi(m^2_h)\equiv \Pi(0)-\Pi(m^2_h)$ in eq.~\eqref{eq:lambdam2H}. For the numerical calculations, the full one-loop value of the Higgs self-energy was used, obtained from the results of ref.~\cite{Martin:2003it}; two loop corrections coming from $\Delta \Pi(m^2_h)$ are expected to be small since they involve scales proportional to $m^2_h\sim \lambda v^2$, so that for small $\lambda<g_i,\lambda_{SH}$ these corrections will be subdominant with respect to the two-loop corrections already included in the effective potential \cite{Degrassi:2012ry}. 
Although eqs.~\eqref{eq:lambdam2H}  were solved numerically using the full two-loop potential as well as the one-loop value of $\Delta \Pi(m^2_h)$, the one-loop solutions for $\lambda,m^2_H$ can be approximated analytically as follows, if one neglects $\Delta \Pi(m^2_h)$ and the contributions to $V^{(1)}$ that depend on  $m^2_H$ and $\lambda$:
\begin{align}
 \nonumber m^2_H=&-\frac{1}{2} {\tilde v}^2 \lambda\!+\frac{1}{16\pi^2}\left[\tilde{v}^2 \left(g_1^4 \left(\frac{9}{400}-\frac{27}{400} \log \left(\frac{m^2_Z(\tilde v)}{v^2}\right)\right)+g_2^4 \left(-\frac{1}{8} 3 \log \left(\frac{m^2_W(\tilde v)}{ v^2}\right)\right.\right.\right.\\
 %%%%
 \nonumber &\left.-\frac{3}{16} \log \left(\frac{m^2_Z(\tilde v)}{v^2}\right)+\frac{3}{16}\right)+g_1^2 g_2^2 \left(\frac{3}{40}-\frac{9}{40} \log \left(\frac{m^2_Z(\tilde v)}{v^2}\right)\right)\\
 %%%
 \nonumber &\left.+\frac{N\lambda _{SH}^2}{2} \left(1-\log \left(\frac{m^2_S(\tilde v)}{v^2}\right)\right)+{y_t}^4 \left(3 \log \left(\frac{{y_t}^2 \tilde{v}^2}{2 v^2}\right)-3\right)\right)\\
 %%%
 &\label{eq:m2H}\left.+N m^2_S \lambda _{SH} \left(1- \log \left(\frac{m^2_S(\tilde v)}{v^2}\right)\right)\right],\\
 %%%%%
 \nonumber \lambda=&\frac{m_h^2}{{\tilde v}^2}\!+\!\frac{1}{16\pi^2}\left[g_1^4 \left(\!-\frac{27}{200}\!  \log \left(\frac{m^2_Z(\tilde v)}{v^2}\right)\!-\!\frac{9}{100}\right)
  \!+\!g_2^2 g_1^2 \left(\!-\frac{9}{20}  \log \left(\frac{m^2_Z(\tilde v)}{v^2}\right)\right.\right.\\
  %%%%%
  \nonumber &\left.-\frac{3}{10}\right)+g_2^4 \left(\!-\frac{3}{4}  \log \left(\frac{m^2_W(\tilde v)}{ v^2}\right)-\frac{3}{8} \log \left(\frac{m^2_Z(\tilde v)}{v^2}\right)-\frac{3}{4}\right)\\
  %%%%
  \nonumber &\left.-N\lambda _{SH}^2  \log \left(\frac{m^2_S(\tilde v)}{v^2}\right)+y_t^4 \left(6 \log \left(\frac{\tilde{v}^2 y_t^2}{v^2}\right)-6 \log (2)\right)\right],
\end{align}
In the expressions above, all couplings should be understood as evaluated at the scale $\mu=v(m_Z)$. Note how $m^2_H(v)$ receives positive contributions proportional to $N\lambda_{SH},N\lambda_{SH}^2$. This ensures that for large enough values of the Higgs-portal coupling, $m^2_H(v)$ can be positive, guaranteeing a concave potential at the origin. Furthermore, $\lambda$ receives negative corrections proportional to $N\lambda_{SH}^2$, and thus $\lambda(v)$ can be negative, which will generate a potential barrier with respect to the origin. The value of $\lambda_{SH}$ needed to get a nonzero barrier will scale approximately as $\frac{1}{\sqrt{N}}$, as follows from setting $\lambda=0$ in eq.~\eqref{eq:m2H}. Perturbative corrections involving the Higgs-portal coupling will in general go as $\lambda_{SH}N\sim \sqrt{N}$, suggesting that models with fewer singlets will be under better perturbative control despite having larger values of $\lambda_{SH}$. The destabilizing effect of a negative quartic $\lambda$ is compensated by the rest of the contributions of the new scalar multiplet to the effective potential. This can be argued from the fact that the contributions from the singlet fields to the effective potential are positive for $m^2_S\geq0,\lambda_{SH}>0$, as is clear from eqs.~\eqref{eq:V1loop}, \eqref{eq:massesh}, or using the arguments of \S~\ref{sec:VCW} linking the large-field behavior of the potential with the running couplings evaluated at a scale proportional to the field itself. Indeed, the potential at large values of the fields can be approximated by $\frac{1}{8}\lambda(\mu=h)h^4$, and given that the beta function of $\lambda$ receives positive contributions proportional to $\lambda_{SH}^2$ (see \S~\ref{app:RG}), the running Higgs quartic coupling will become positive for large scales, ensuring stability.

\begin{figure}[h!]\centering
  \includegraphics[scale=.8]{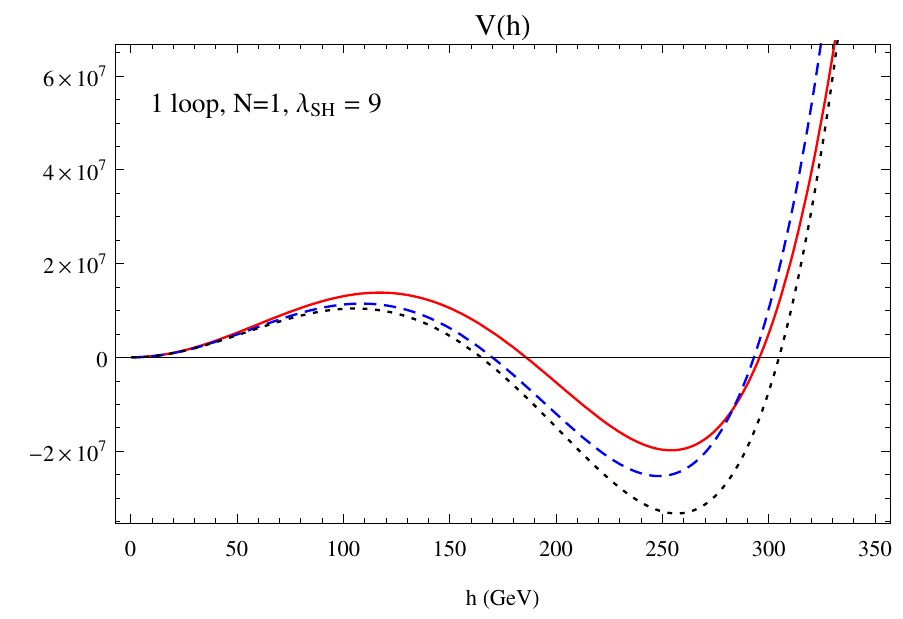}
    \includegraphics[scale=.8]{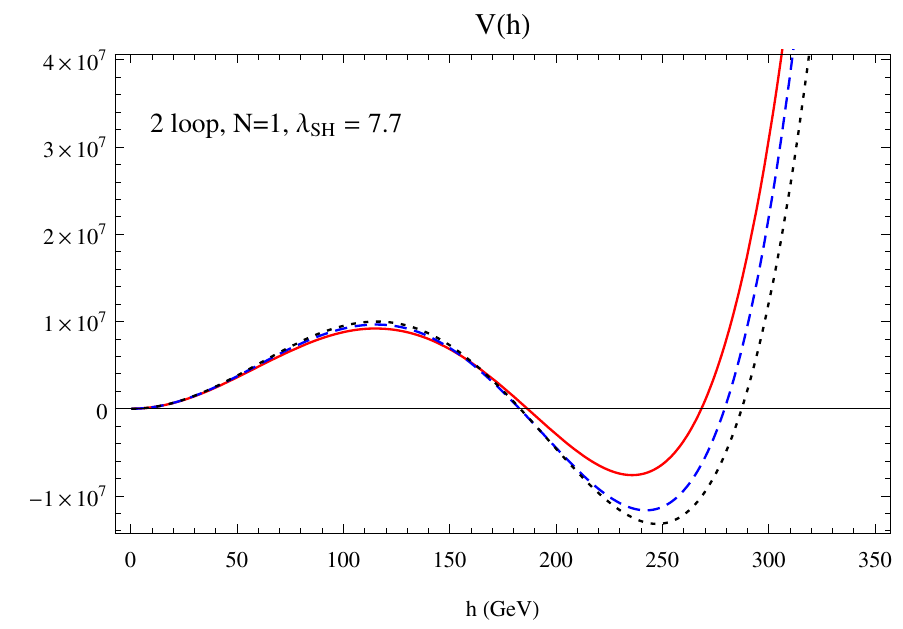}
     \includegraphics[scale=.8]{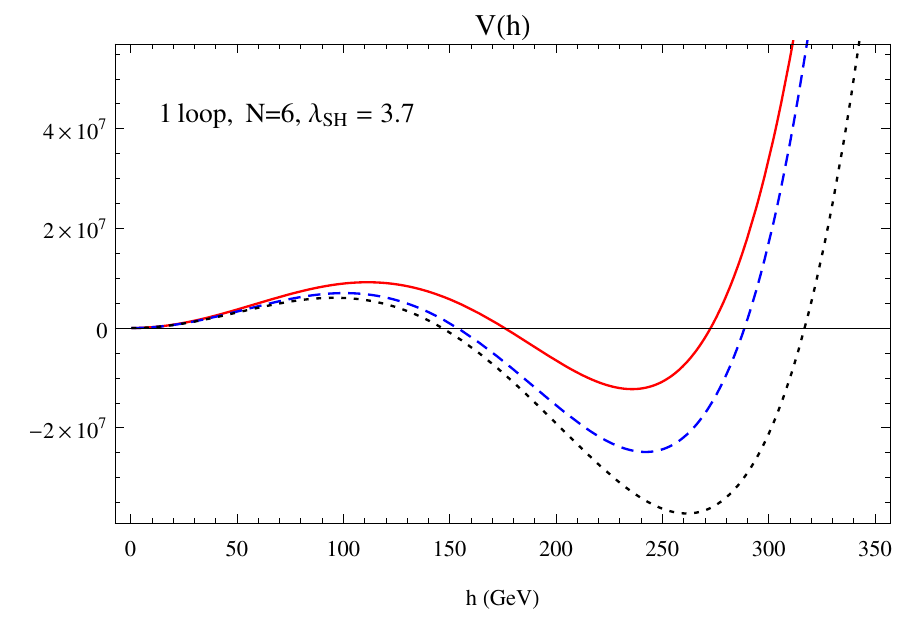}
    \includegraphics[scale=.8]{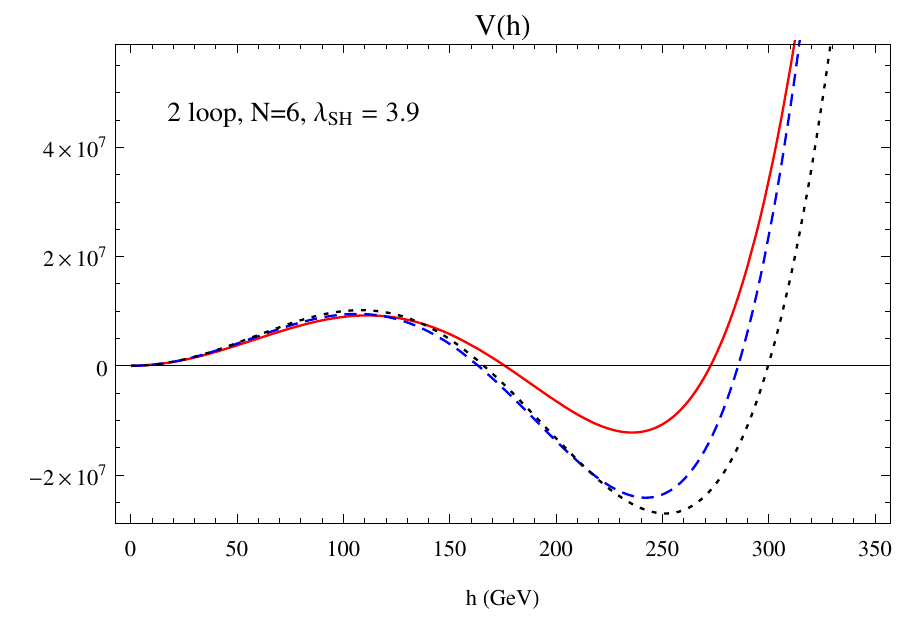}
     \includegraphics[scale=.8]{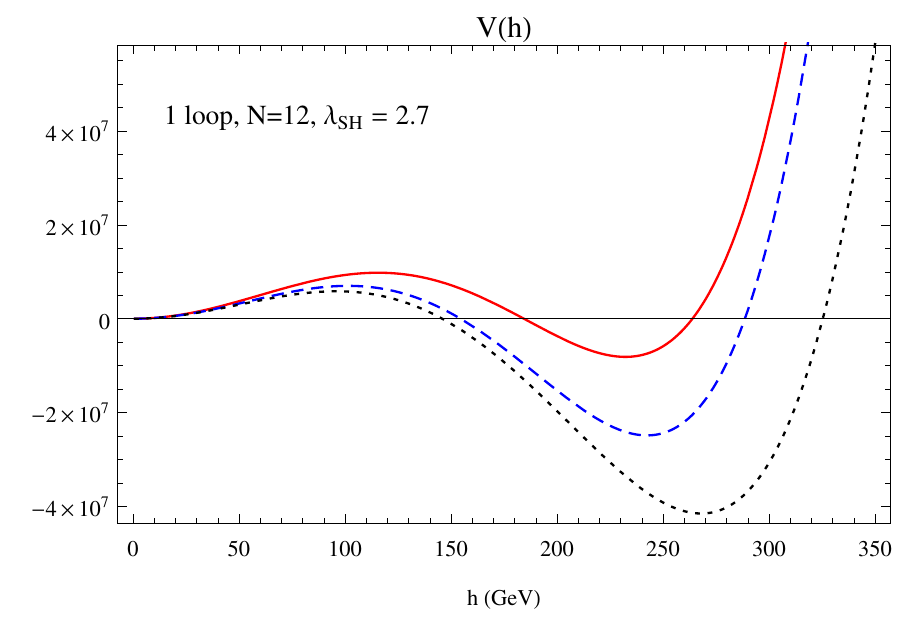}
    \includegraphics[scale=.8]{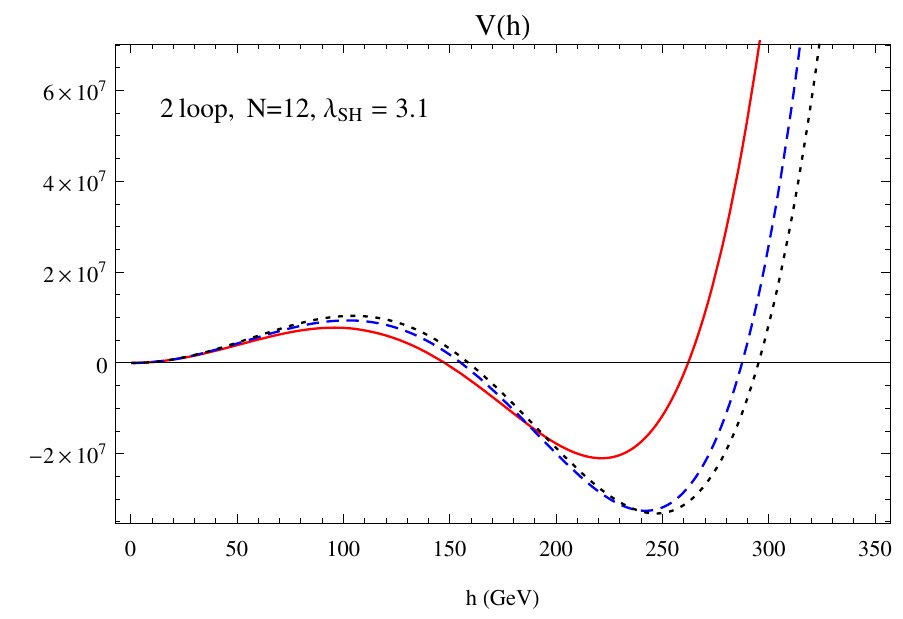}
\caption{\label{fig:V}Effective potentials in GeV units at one (left) and two-loops (right), for $N=1$ (top), $N=6$ (middle row), $N=12$ (bottom), for different values of $\lambda_{SH}$ yielding a large barrier. $\lambda_S$ and $m^2_S$ were fixed at $0.1$ and $10^4 \text{ GeV}^2$, respectively. The behavior under rescalings is illustrated by choosing $\mu=175$ GeV (solid red), $\mu=246$ GeV (blue, dashed), and a field-dependent choice $\mu=\exp(-h^2/3\cdot 10^4)+h$ (black, dotted). Note the improved scaling behavior in the two-loop case.}
\end{figure}

\begin{figure}[h!]\centering
  \includegraphics[scale=1.2]{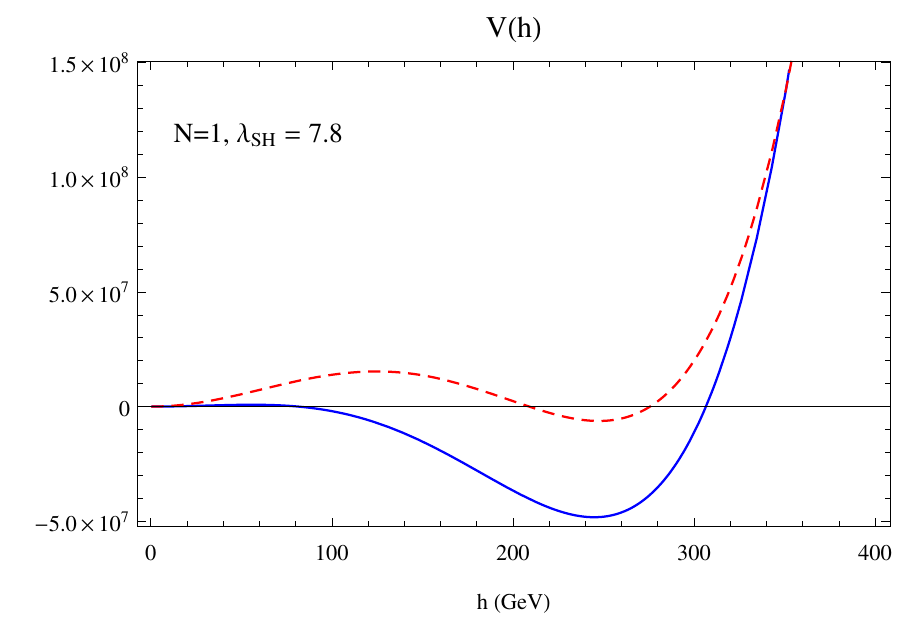}
\caption{\label{fig:V12}Effective potentials in GeV units at one (blue) and two-loops (dashed red), for $N=1, \lambda_S=0.1, \lambda_{SH}=7.8, m^2_S=10^4 \text{ GeV}^2, \mu=\exp(-h^2/3\cdot 10^4)+h$.}
\end{figure}

To illustrate the previous observations, fig.~\ref{fig:V} shows the effective potential at one and two-loops, for $N=1,\,N=6$ and $N=12$, choosing values of the couplings leading to a sizable barrier, and for different choices of the renormalization scale. The vacuum energy was subtracted as commented before, allowing to use field-dependent choices of scale. It is to be noted how the behavior under changes of scale is significantly improved at two-loops, despite the large values of the coupling $\lambda_{SH}$. This is a strong indication for a well-behaved perturbative expansion. Particularly, the appearance of a zero-temperature barrier is confirmed beyond any reasonable doubt, given the remarkable scale-independence of the barrier height at two-loops. Still, the value of the Higgs-portal coupling needed to get a given barrier size does change between one and two-loops, as is clear from fig.~\ref{fig:V} or from the direct comparison between one and two-loop results shown in fig.~\ref{fig:V12}. Also, the results validate the previous arguments concerning the strength of the couplings and the reliability of perturbative computations: although  a greater number of singlets implies that a smaller value of $\lambda_{SH}$ suffices to generate a barrier, this does not necessarily imply a more reliable perturbative behavior. As illustrated by fig.~\ref{fig:V}, the scaling behavior is better in the case of one singlet with a coupling $\lambda_{SH}\sim8$ than in the case of twelve singlets with a coupling $\lambda_{SH}\sim3$. 

\begin{figure}[h!]\centering
  \includegraphics[scale=.65]{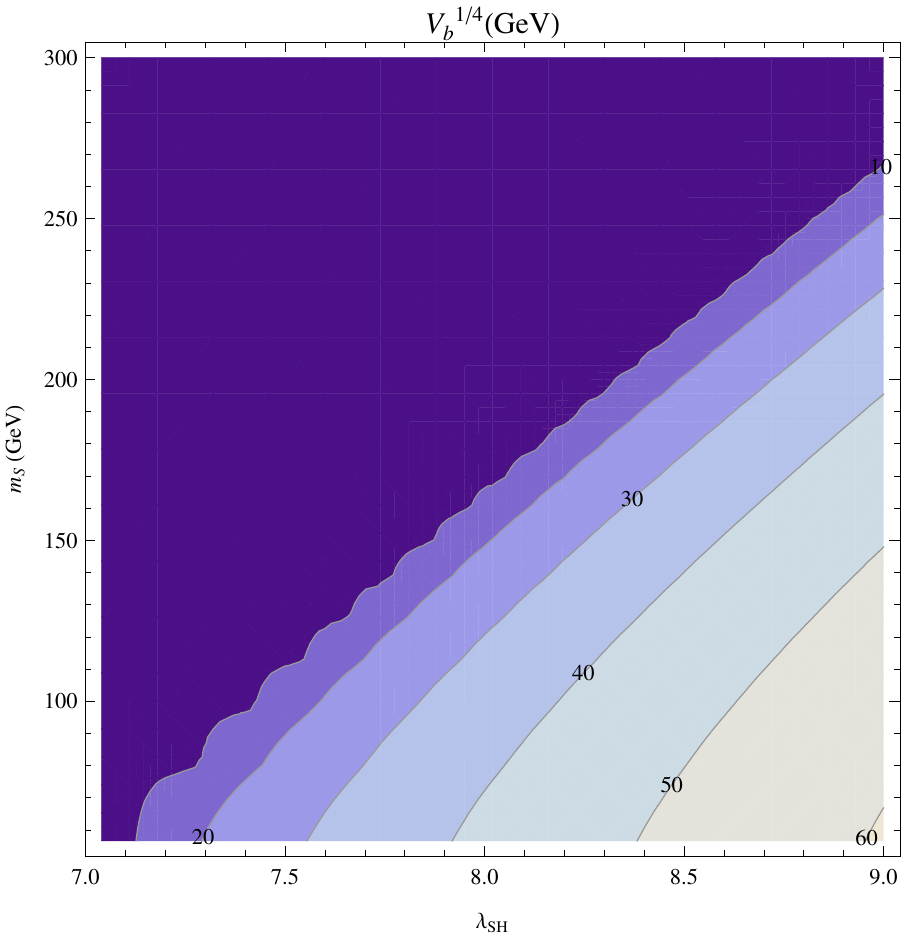}
    \includegraphics[scale=.65]{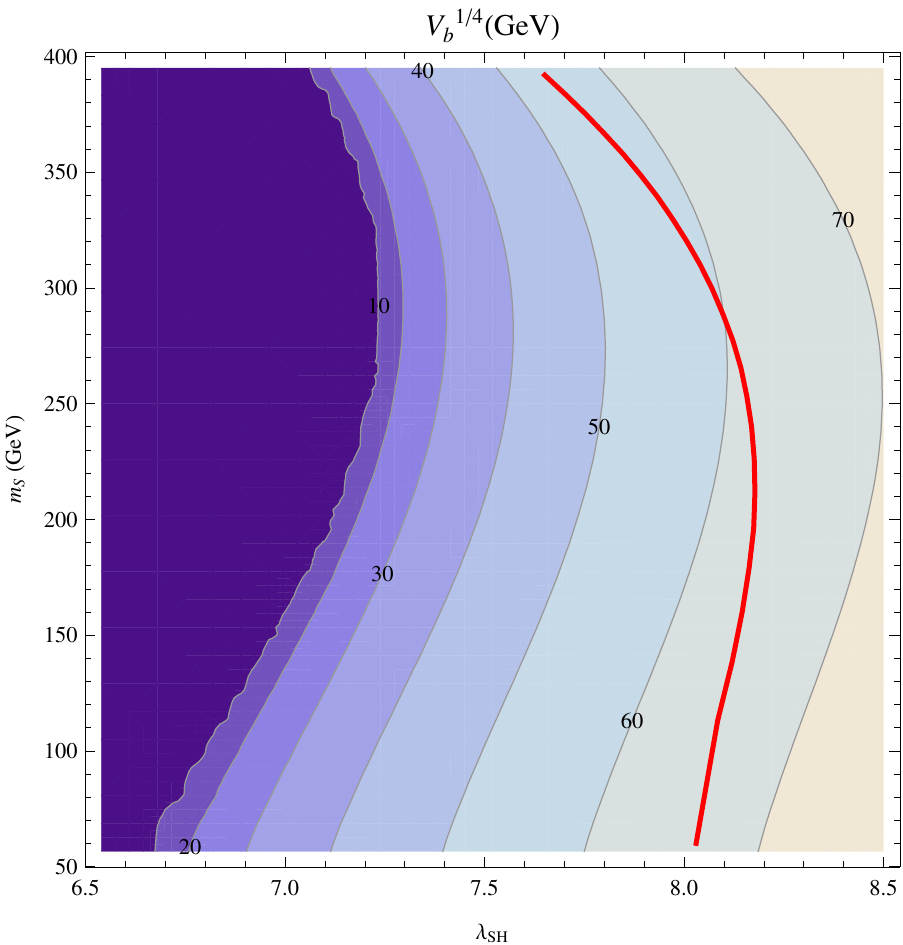}
      \includegraphics[scale=.65]{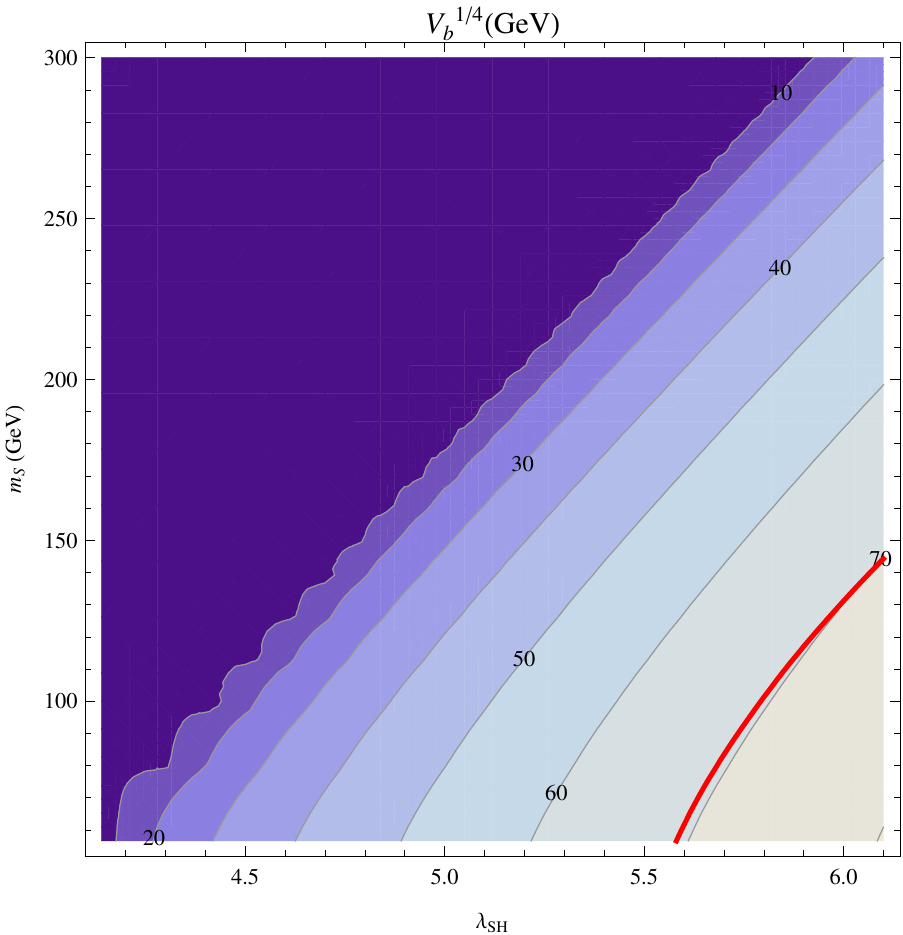}
    \includegraphics[scale=.65]{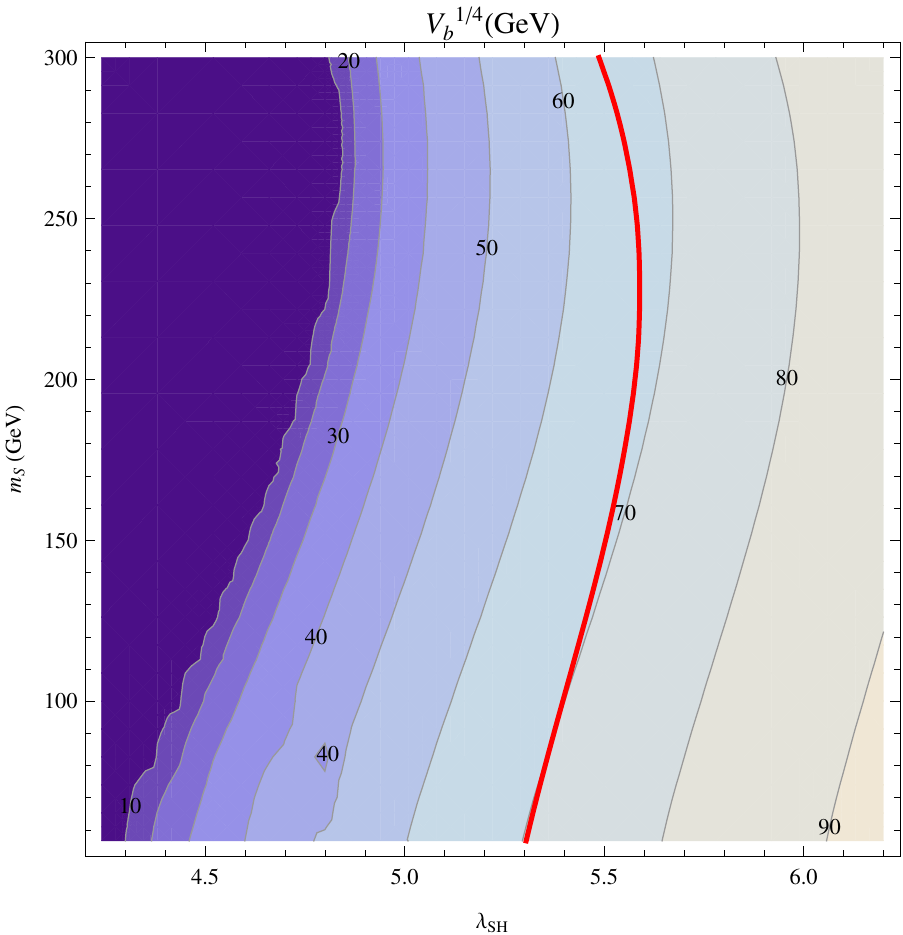}
      \includegraphics[scale=.65]{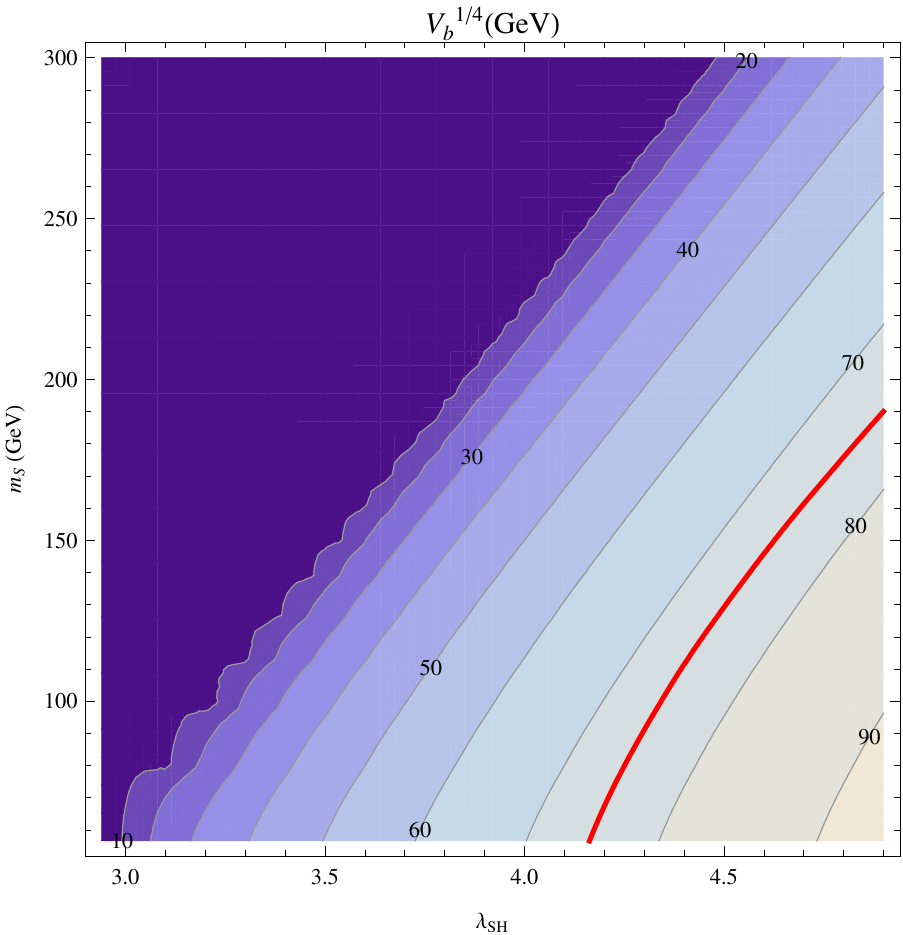}
    \includegraphics[scale=.65]{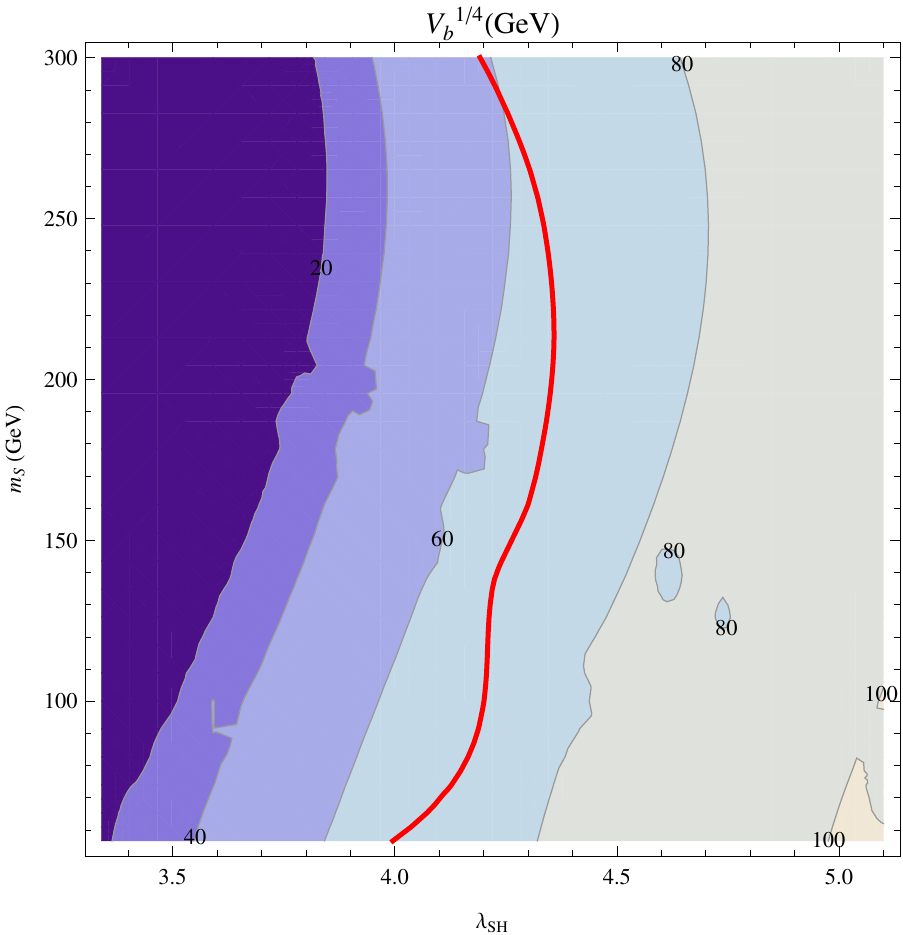}
\caption{\label{fig:Vbarrier}Quartic root of the barrier height in GeV units at one (left) and two-loops (right), for $N=1$ (top), $N=3$ (middle row), $N=6$ (bottom), in terms of $\lambda_{SH}$ and $m_S$. $\lambda_S$ was fixed at 0.1. Points to the right of the red line have an electroweak vacuum with energy greater than $V(h=0)$.  }
\end{figure}

The height of the barrier is shown in fig.~\ref{fig:Vbarrier} for $N=1,\,N=3,\,N=6$, for fixed $\lambda_{S}=0.1$, in terms of $m^2_S$ and $\lambda_{SH}$. As the Higgs-portal coupling grows, the energy of the electroweak vacuum increases, eventually becoming higher than the potential at the origin; this happens to the right of the red lines in the figure. As mentioned in the introduction,  Coleman-Weinberg realizations of electroweak symmetry breaking arise as limiting cases  in which the mass parameters of the theory are zero. In order to identify these scenarios, $m^2_S$ was fixed to be zero at the scale of the Higgs VEV and, for fixed $\lambda_S=0.1$, $\lambda_{SH}$ was probed in search of scenarios yielding $m^2_H=0$ after imposing the conditions of eq.~\eqref{eq:lambdam2H}. At least for the range of values of $m^2_S$ that was explored, the corresponding values of $\lambda_{SH}=\lambda_{SH}^{CW}$ for each $N$ are the minimum values needed to have a nonzero barrier, as is clear from fig.~\ref{fig:Vbarrier}, in which the Coleman-Weinberg scenarios would sit at the bottom-left corner of the graphs. The resulting Coleman-Weinberg values of $\lambda_{SH}$  were checked to be largely insensitive to the value of $\lambda_S$, and are summarized in the following table:
\begin{minipage}{\linewidth}
\begin{center}
\begin{tabular}{c|c|c|c|c}
N & 1 & 3 & 6 & 12\\
\hline
$\lambda(v)_{SH}^{CW,\text{1 loop}}$ & 6.81 & 3.95 & 2.81 & 2.00\\
\hline
$\lambda(v)_{SH}^{CW,\text{2 loop}}$ & 6.89 & 4.10 & 2.93 & 2.10
\end{tabular}
\captionof{table}{Parameter values for Coleman-Weinberg scenarios.}\label{table:CW}
\end{center}
\end{minipage}

\begin{figure}[h!]\centering
  \includegraphics[scale=1.1]{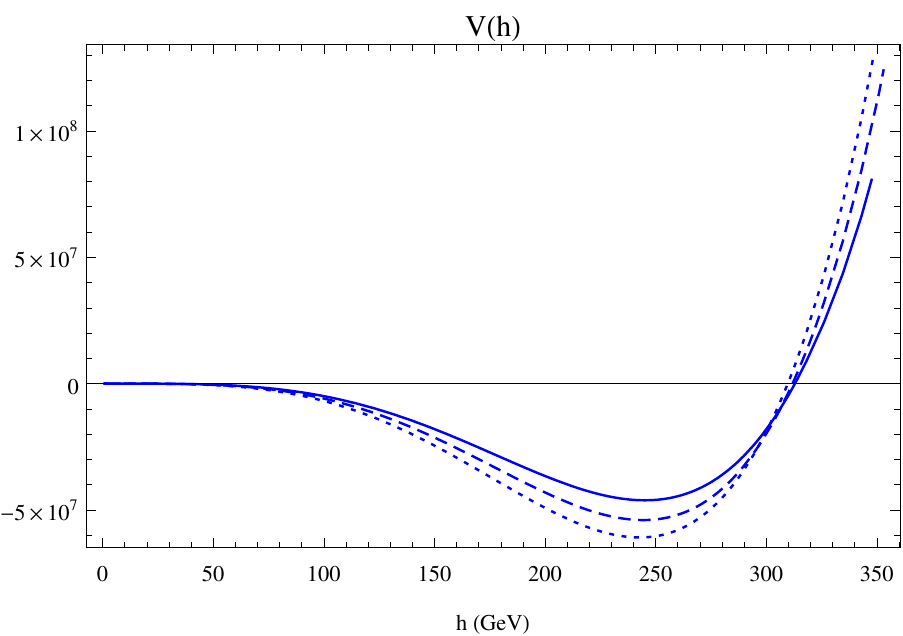}
\caption{\label{fig:VCW}Two-loop effective potentials in the Coleman-Weinberg scenarios for $N=1$ (solid), $N=3$ (dashed), and $N=6$ (dotted).}
\end{figure}

The resulting Coleman-Weinberg potentials are illustrated in fig.~\ref{fig:VCW} for $N=1,3,6$.

In ref.~\cite{Espinosa:2007qk}, which studied models with twelve real scalars (corresponding to $N=6$ in this paper), a zero barrier height with $m_h=125$ was achieved for $\lambda_{SH}\sim2.7$,  close to the one-loop value of table~\ref{table:CW}\footnote{The coupling $\zeta$ of ref.~\cite{Espinosa:2007qk} is related to the Higgs-portal coupling $\lambda_{SH}$ in this article as $\lambda_{SH}=2\zeta^2$.}. Two-loop corrections in the value of $\lambda^{CW}_{SH}$ for $N=6$ are of the order of 4\% --as opposed to 1\% for $N=1$, again indicating a better perturbative behavior for smaller $N$.

To end the discussion about the zero-temperature effective potential, it is important to note that in these models with a strong Higgs-portal coupling, the  quartic couplings typically reach a one-loop Landau pole near the TeV scale, while at two-loops the theory seems to flow to a strongly coupled fixed point with positive couplings. Either way, this signals that perturbation theory is likely to break down not far from the TeV scale. This behavior can be ameliorated by charging the singlets under a hidden gauge group, whose effects on the running quartics can keep them perturbative until the Planck scale \cite{Dermisek:2013pta}. At the energies near the electroweak scale relevant for the issues tackled in this paper, such modifications of the model are not expected to yield significant differences, since as it was seen perturbation theory is already well-behaved.

 %%%%%%%%%%%%%%%%%%%%%%%%%%%%
 %%%%%%%%%%%%%%%%%%%%%%%%%%%%
 %%%%%%%%%%%%%%%%%%%%%%%%%%%%
 %%%%%%%%%%%%%%%%%%%%%%%%%%%%
 
 \section{Finite-temperature effects \label{sec:FT}}
 
 Thermal effects can be evaluated from finite-temperature contributions to the effective potential, including a Daisy resummation of the large infrared corrections due to bosonic fields; see for example refs.~\cite{Quiros:1999jp,Kapusta:2006pm}. The thermal corrections can be expressed as
 \begin{align}
 \label{eq:VT}
  \Delta V(h,T)=\frac{T^4}{2\pi^2}\left[\sum_B J_B\left(\frac{m^2_B(h)}{T^2}\right)-\sum_F J_F\left(\frac{m^2_F(h)}{T^2}\right)\right]+\Delta V^{(2)},
 \end{align}
 where the index $B$ runs over bosonic degrees of freedom --vectors and real scalars-- and the index $F$ over Weyl fermions. The functions $J_B$ and $J_F$ encoding the one-loop corrections are given by
 \begin{align*}
  J_B(x)=\int_0^\infty dy\, y^2\log\left[1-\exp(-\sqrt{x^2+y^2})\right],\\
  %%%%
    J_F(x)=\int_0^\infty dy\, y^2\log\left[1+\exp(-\sqrt{x^2+y^2})\right],
 \end{align*}
and their large temperature expansions are given by
\begin{align}
\label{eq:exp}
 T^4 J_B\left(\frac{m^2}{T^2}\right)=&-\frac{\pi^4 T^4}{45}+\frac{\pi^2m^2 T^2}{12}-\frac{T\pi (m^2)^{3/2}}{6}-\frac{(m^4)}{32}\log\frac{m^2}{a_b T^2},\\
 %%%%
 \nonumber T^4 J_F\left(\frac{m^2}{T^2}\right)=&\frac{7\pi^4 T^4}{360}-\frac{\pi^2m^2 T^2}{24}-\frac{(m^4)}{32}\log\frac{m^2}{a_f T^2}, 
\end{align}
with $a_b=16\pi^2e^{3/2-2\gamma_E}$ and $a_f=\pi^2e^{3/2-2\gamma_E}$, $\gamma_E$ being the Euler constant. The function $J_B$ has a nonanalytic behavior  due to an infrared singularity in the limit of zero mass, as evidenced by the $(m^2)^{3/2}$ power in eq.~\eqref{eq:exp}. As a consequence of this, for small bosonic masses it is important to resum diagrams with similar infrared singularities, the dominant contributions coming from the so-called ring or Daisy diagrams. The resummation can be seen to be equivalent to substituting the tree-level, field dependent masses with their one-loop, finite-temperature  values \cite{Kapusta:2006pm,Arnold:1992rz}. The relevant fields are in this case the longitudinal gauge bosons $W^\pm_L, Z_L, \gamma_L$ and the Higgs scalars (given the large Higgs-portal couplings, the singlets are heavy, making the resummation unnecessary). The thermally corrected masses of the longitudinal gauge bosons are given by \cite{Espinosa:1993bs}
\begin{align*}
 {\tilde m}^2_{W^\pm_L}=&m^2_W(h)+\frac{11}{6}g^2 T^2,\\
 %%%%
 {\tilde m}^2_{Z_L}=&\frac{1}{2}\left[m^2_Z(h)+\frac{11}{6}\frac{g^2}{\cos^2\theta_W} T^2+\Delta(h,T)\right],\end{align*}\begin{align*}
 %%%
 {\tilde m}^2_{\gamma_L}=&\frac{1}{2}\left[m^2_Z(h)+\frac{11}{6}\frac{g^2}{\cos^2\theta_W} T^2-\Delta(h,T)\right],\\
 %%%%%%%
 \Delta(h,T)=&m^4_Z(h)+\frac{11}{3}\frac{\cos^2 2\theta_W}{\cos^2\theta_W}T^2\left[m^2_Z(h)+\frac{11}{12}\frac{g^2}{\cos^2\theta_W}T^2\right],
\end{align*}
where $\theta_W$ designates the Weinberg angle, and the field-dependent masses $m^2_a(h))$ are given at the end of eq.~\eqref{eq:massesh}. Regarding the Higgs masses, the fact that these might be negative at tree-level given the negative values of $\lambda$ typical in these scenarios is problematic for the numerical evaluation of  thermal corrections, so that it is important to resum not only finite-temperature contributions to the masses, but positive zero-temperature corrections as well. Writing the Higgs doublet as $H=(\frac{1}{\sqrt{2}}(h^+_r+ih^+_i),\frac{1}{\sqrt{2}}(h+i\chi))^\intercal$, the one-loop corrected masses for the fields $h$ and $G_i=\chi, h^+_r,h^+_i$ are, including the effects of the quartic couplings and the top Yukawa,
\begin{align}
\nonumber \tilde m^2_h(h)=& m^2_{h}(h)+T^2\left(\frac{\lambda}{4}+\frac{N\lambda_{SH}}{12}+\frac{y_t^2}{4}\right)+\frac{1}{16\pi^2}\left[\lambda  m^2_H \left(\frac{3}{2} \log \left(\frac{ m^2_{G}(h)}{\mu^2}\right)+\frac{3}{2} \log \left(\frac{ m^2_{h}(h)}{\mu^2}\right)\right.\right.\\
 %%%%
 \nonumber&\left.-3\right)+h^2 \left(\lambda ^2 \left(\frac{9}{4} \log \left(\frac{ m^2_{G}(h)}{\mu^2}\right)\!+\frac{27}{4} \log \left(\frac{ m^2_{h}(h)}{\mu^2}\right)\!-\!3\right)+N\lambda _{\text{SH}}^2 \left(\frac{3}{2} \log \left(\frac{m^2_{S}(h)}{\mu^2}\right)\right.\right.\\
 %%%%
 \nonumber&\left.\left.\left.-\frac{1}{2}\right)+y_t^4 \left(3-9 \log \left(\frac{h^2 {y_t}^2}{2 \mu ^2}\right)\right)\right)+N m^2_S \lambda _{\text{SH}} \left( \log \left(\frac{m^2_{S}(h)}{\mu^2}\right)-1\right)\right],\\
 %%%%
\nonumber \tilde m^2_{G_i}(h)=& m^2_{G_i}(h)+T^2\left(\frac{\lambda}{4}+\frac{N\lambda_{SH}}{12}+\frac{y_t^2}{4}\right)
+\frac{1}{16\pi^2}\left[\frac{5\lambda}{2} m^2_{G_i}(h)\left(\log\frac{ m^2_{G_i}(h)}{\mu^2}-1\right)\right.\\
%%%
\nonumber&+\frac{\lambda}{2} m^2_{h}(h)\left(\log\frac{ m^2_{h}(h)}{\mu^2}-1\right)+N\lambda_{SH} m^2_{S}(h)\left(\log\frac{ m^2_{S}(h)}{\mu^2}-1\right)\\
%%%%
&\label{eq:mhresummed}\left.+3h^2y_t^4\left(1- \log \left(\frac{h^2 {y_t}^2}{2 \mu ^2}\right)\right)\right].
\end{align}
The one-loop mass $\tilde m^2_h(h)$ was obtained from the one-loop effective potential for $h$, while the expression for $m^2_{G_i}(h)$ follows from the two-point functions of the Goldstone fields in the background of $h$, including the leading temperature corrections.

Going back to eq.~\eqref{eq:VT}, the last piece represents two-loop finite-temperature corrections.  In these scenarios with a large Higgs-portal coupling $\lambda_{SH}$, the dominant contributions come from the scalar diagrams in fig.~\eqref{fig:scalardiagrams}, for which formulae are given in appendix \ref{app:diagrams} in terms of integrals that can be evaluated numerically for positive values of the masses (hence the need to resum positive corrections in the case of negative tree-level masses).
\begin{figure}[h!]\centering
  \includegraphics[scale=.45]{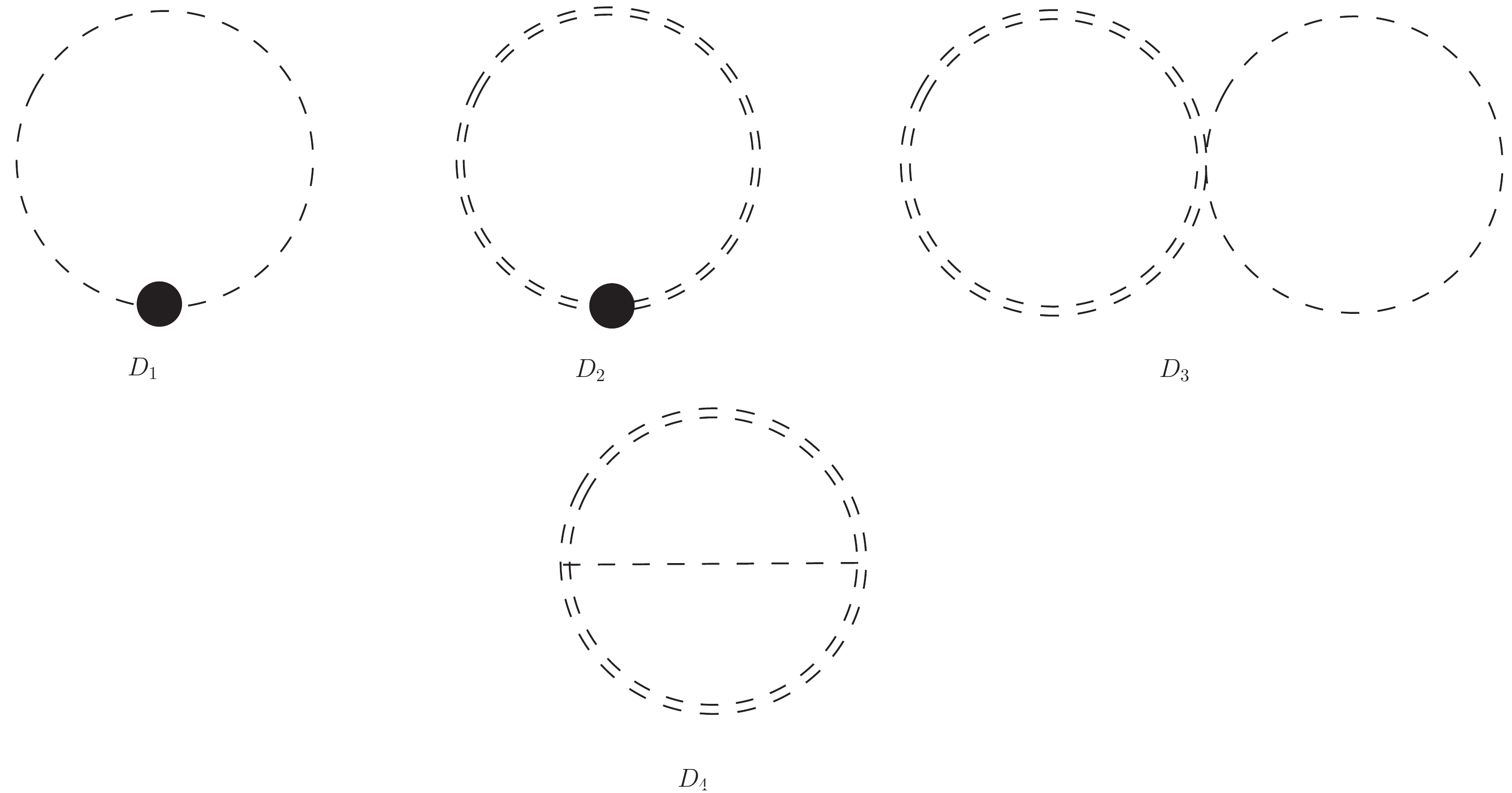}
\caption{\label{fig:scalardiagrams}Two-loop diagrams involving the Higgs-portal coupling and a nonzero background for the Higgs field. Higgs propagators are denoted with single lines, and singlet propagators with double lines. The dots denote one-loop counterterms.}
\end{figure}

In order to gain a qualitative understanding of temperature effects, let's recall that in the Standard Model, in which the symmetry breaking is generated by a negative $m^2_H$ term in the Higgs potential, the symmetry is restored at high temperatures because the thermal corrections include positive quadratic terms, as is clear from the expansions of $J_B, J_F$ in equation \eqref{eq:exp} for $m^2\sim h^2$. In the scenarios which are the focus of this paper, $m^2_H$ is positive and the electroweak vacuum is caused instead by a negative Higgs quartic coupling. Nevertheless, symmetry will be restored again at high temperatures, as is clear from the temperature dependence of the $m^4$ terms in the expansions of eq.~\eqref{eq:exp}, which give positive contributions for the Higgs quartic coupling at high temperature. It should be noted however that the high-temperature expansions in eq.~\eqref{eq:exp}, or the corresponding two-loop expansions given in ref.~\cite{Arnold:1992rz}, should not be trusted  for field values around the electroweak scale and temperatures near the critical one, because $T_c$ turns out to be smaller than $v$. For these reason, thermal corrections were evaluated by computing the functions $J_B$ and $J_F$ numerically and by using the two-loop numerical integrals of \S~\ref{app:diagrams}.

\begin{figure}[h!]\centering
  \includegraphics[scale=.85]{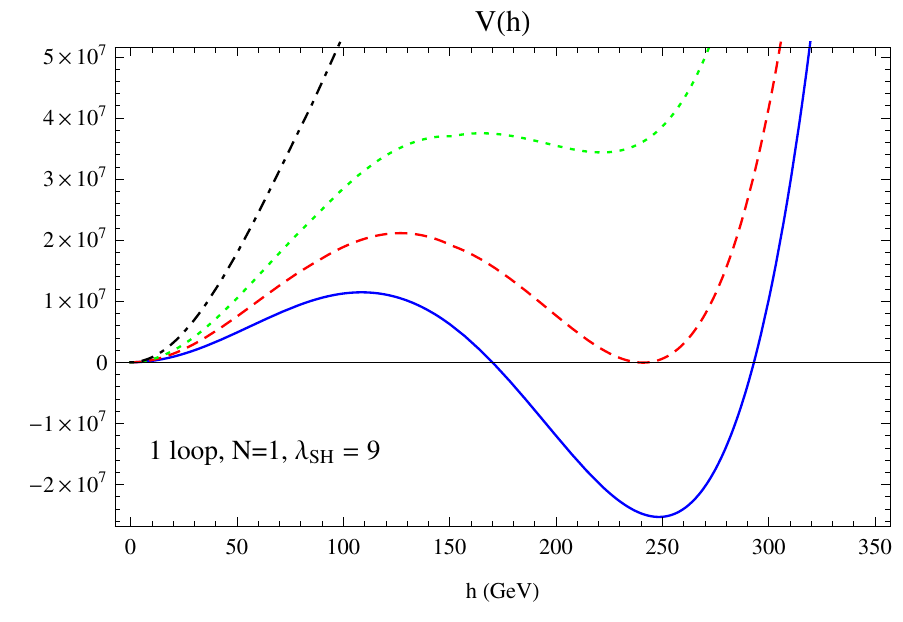}
  \includegraphics[scale=.85]{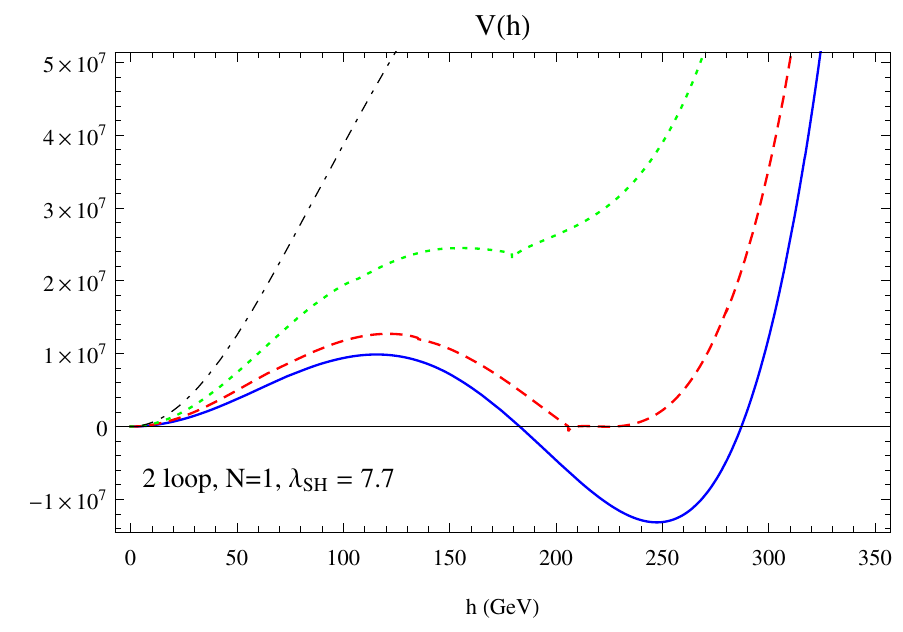}
   \includegraphics[scale=.85]{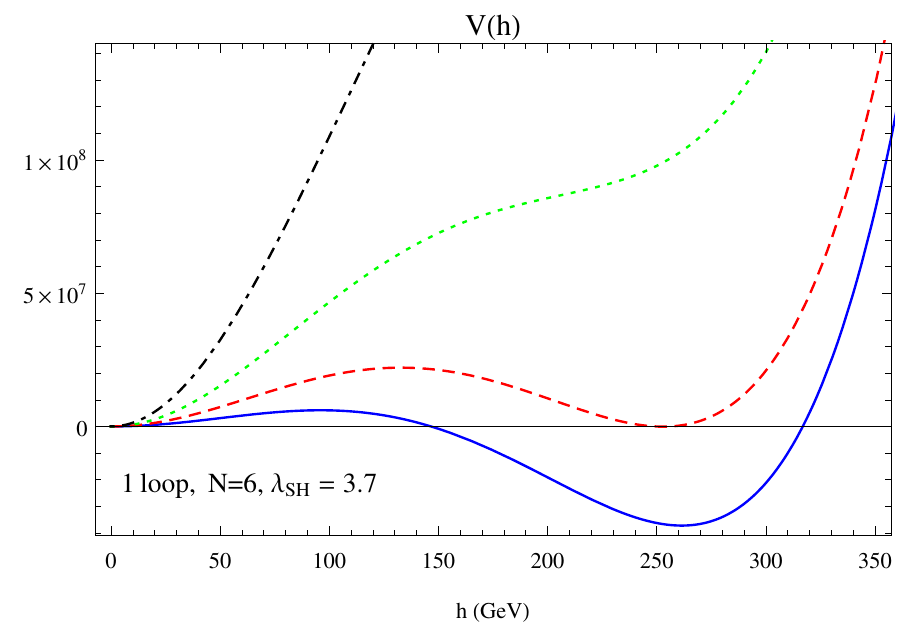}
  \includegraphics[scale=.85]{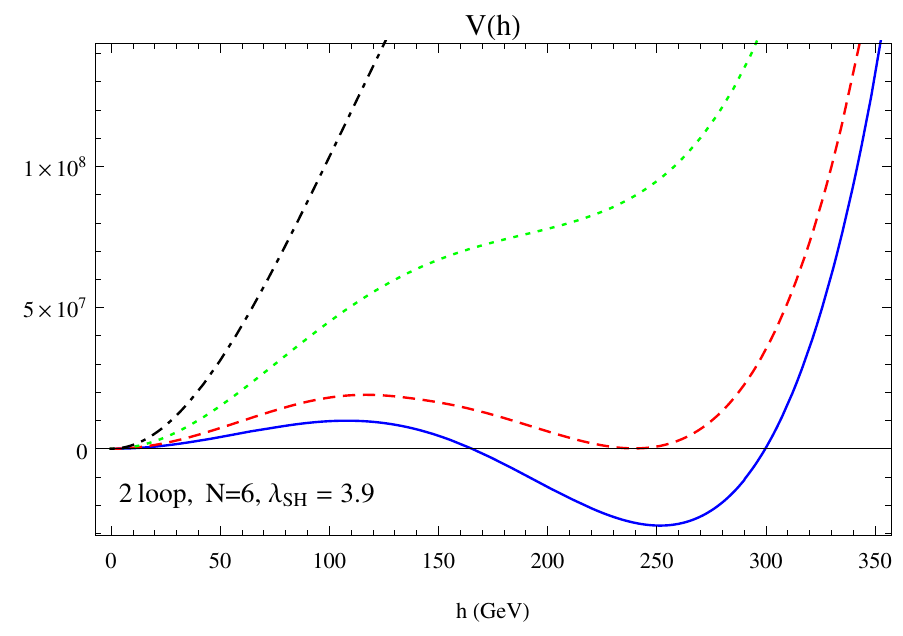}
\caption{\label{fig:VT}Effective potential for $N=1$ (top) and $N=6$ (bottom), at one (left) and two-loops (right), for the same choices of parameters as the corresponding graphs in fig.~\ref{fig:V}. The choices of temperatures are: zero (solid blue), critical temperature (dashed red, from top to bottom and left to right: $74.5, 67.2, 63.2$ and $64.5$ GeV), $100$ GeV (dotted green) and $140$ GeV (dash-dotted black).} 
\end{figure}

Fig.~\ref{fig:VT} shows the resulting behavior of the effective potential under changes of temperature, at one and two-loops, for the $N=1$ and $N=6$ scenarios of fig.~\ref{fig:V}. The cusps in the two-loop case for $N=1$ are due to the field-dependent masses in the Higgs multiplet passing from negative to positive values despite the use of the one-loop resummed masses of eq.~\eqref{eq:mhresummed}. Negative masses were substituted by zero inside the thermal integrals in \S~\ref{app:diagrams}, introducing an artificial cut which nevertheless does not yield a big effect. 

\begin{figure}[h!]\centering
  \includegraphics[scale=.67]{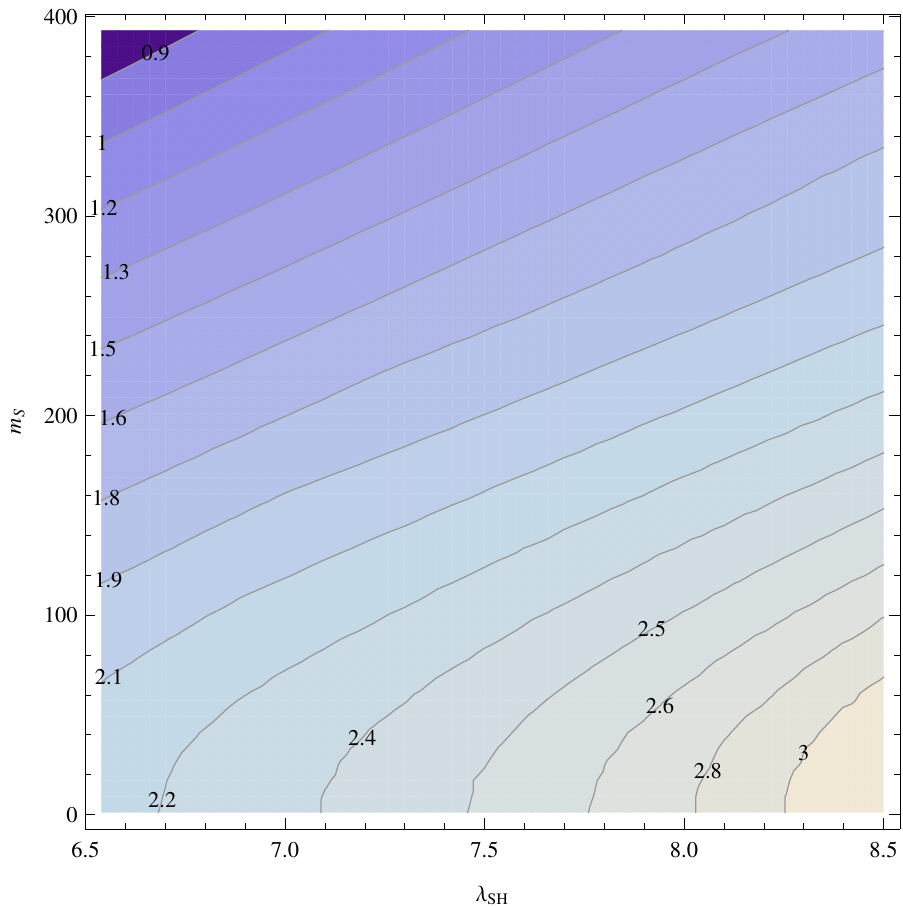}
     \includegraphics[scale=.67]{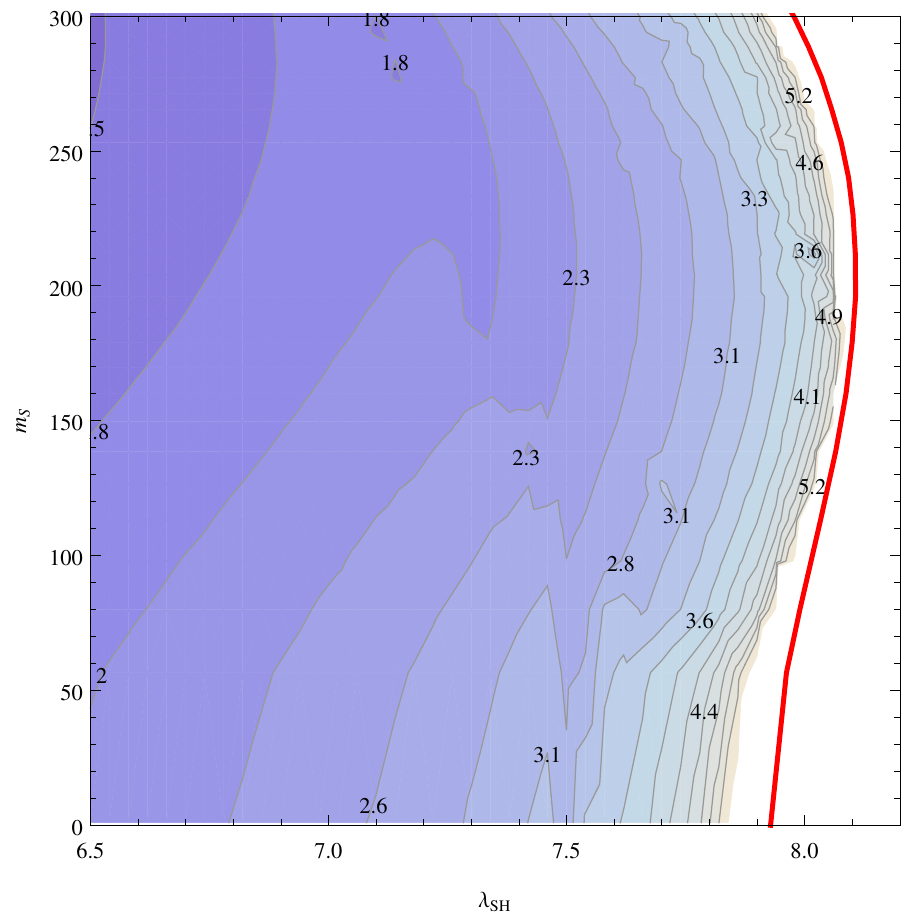}
      \includegraphics[scale=.67]{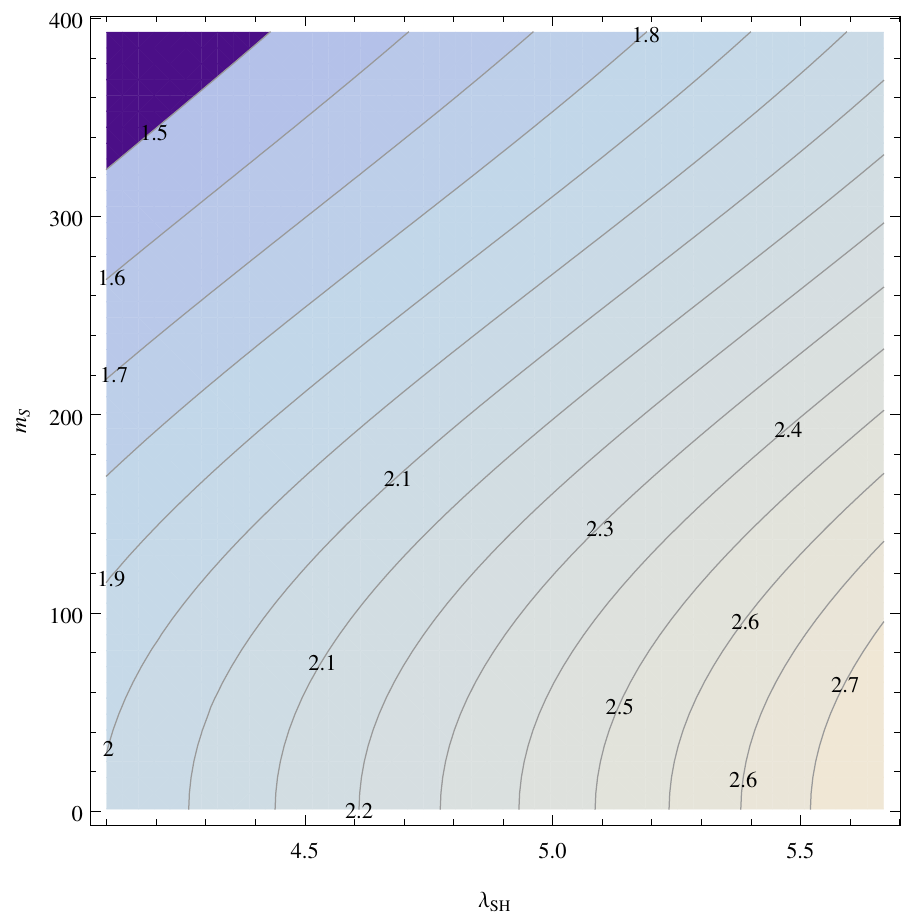}
    \includegraphics[scale=.67]{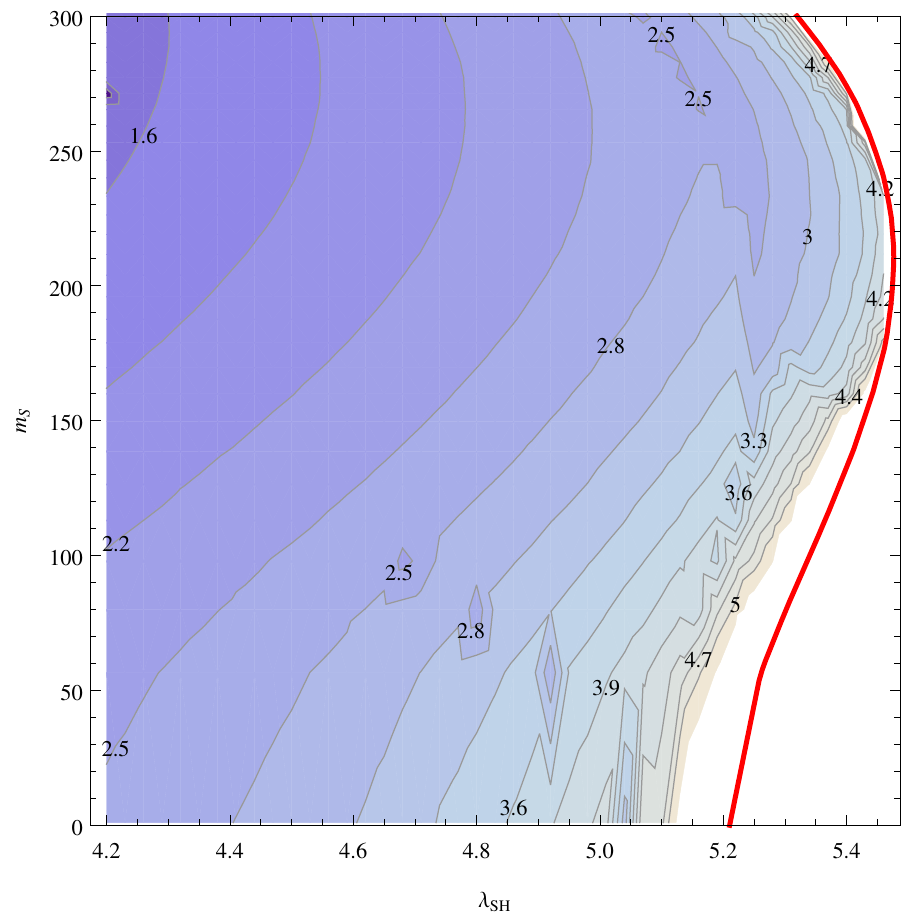}
      \includegraphics[scale=.67]{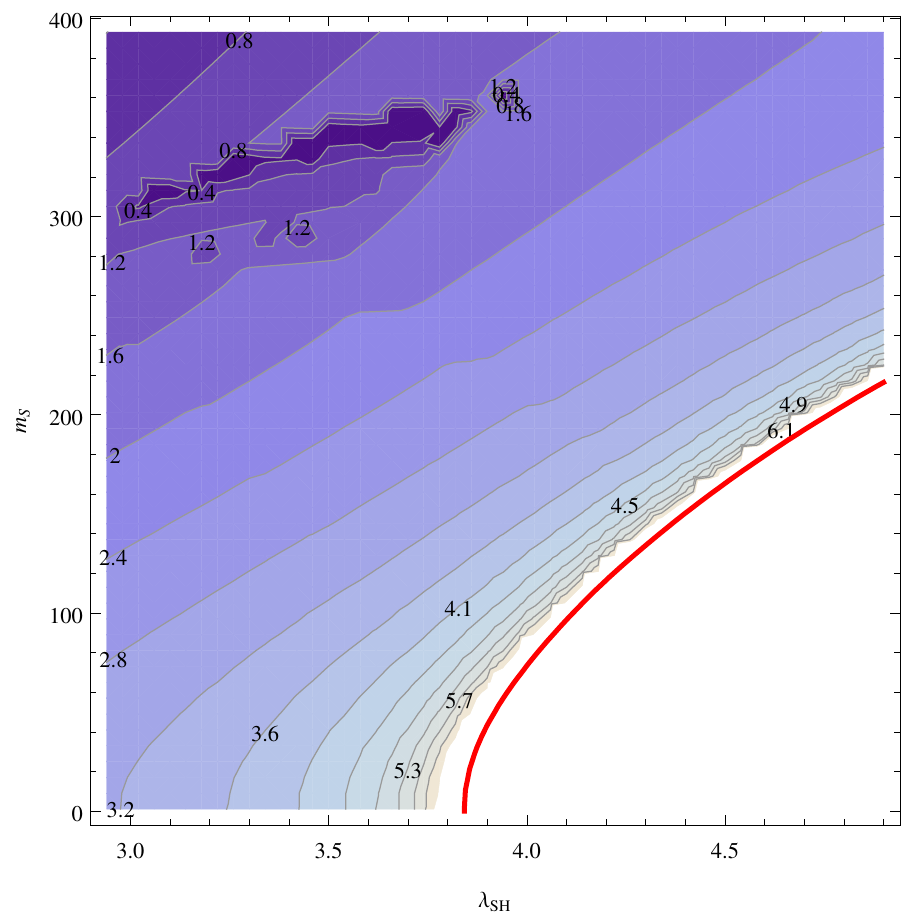}
    \includegraphics[scale=.67]{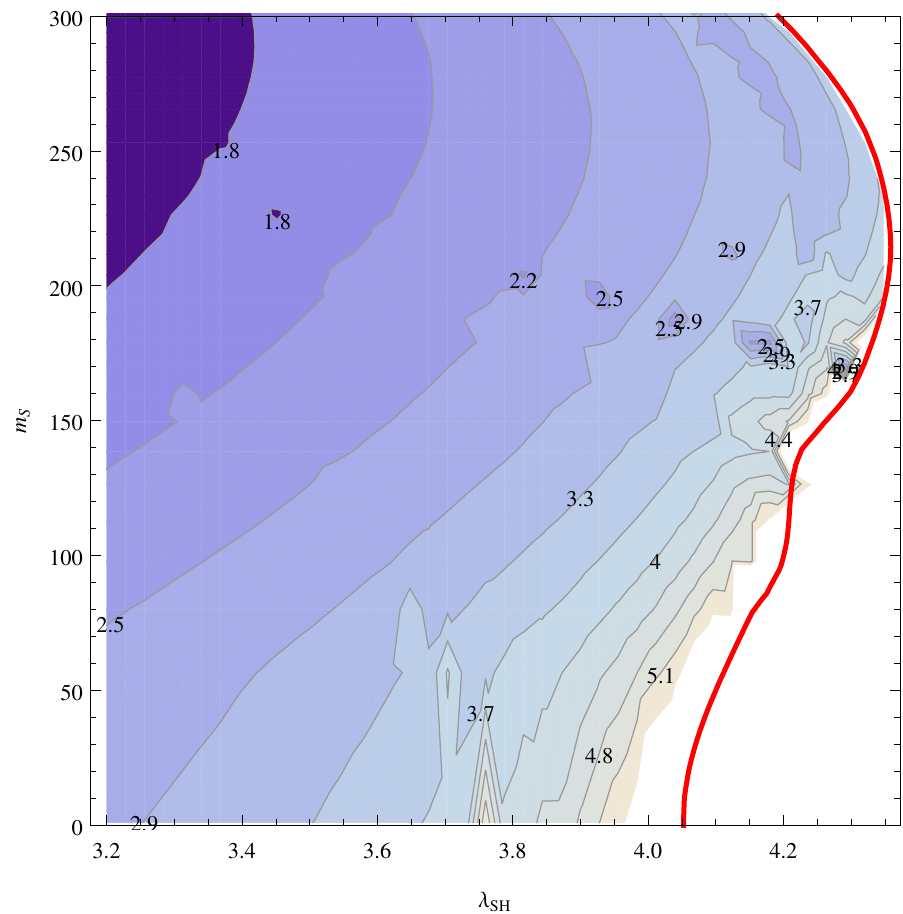}
\caption{\label{fig:vcTc}Strength of the first-order phase transition at one (left) and two-loops (right), for $N=1$ (top), $N=3$ (middle row), $N=6$ (bottom), in terms of $\lambda_{SH}$ and $m_S$. $\lambda_S$ was fixed at 0.1. The red lines correspond to $\frac{v_c}{T_c}=\infty$, and to their right the electroweak vacuum lies above the origin at zero temperature.  }
\end{figure}

As mentioned in the introduction, part of the interest of the scenarios with nonzero potential barriers at zero temperature lies in that they achieve strong first-order electroweak phase transitions. As is clear from the previous discussion, at high temperatures the symmetry is restored and the vacuum lies at the origin, $h=0$. As the temperature is lowered, the new vacuum makes its appearance, separated from the origin by a barrier and with an energy that decreases as the temperature is lowered, until it becomes degenerate with the energy at the origin at a critical temperature $T_c$. The phase transition will be strong first-order if the VEV of the new vacuum at the critical temperature $T_c$ satisfies $v_c/T_c\gtrsim1$, which ensures the suppression of baryon violation processes in the broken phase and thus the survival of any baryon asymmetry generated during the nucleation of bubbles of the true vacuum \cite{Bochkarev:1987wf,Quiros:1999jp}. Fig.~\ref{fig:vcTc} shows contour maps of the values of $v_c/T_c$ for the models studied in this paper. Note how a strong first-order transition is guaranteed even in the Coleman-Weinberg scenarios with zero $m^2_S$ and $\lambda_{SH}$ given by table~\ref{table:CW}. The strength of the phase transition reaches infinity when the zero temperature Higgs vacuum becomes degenerate with the origin, as in the red lines of figs.~\ref{fig:Vbarrier} and ~\ref{fig:vcTc}.

 %%%%%%%%%%%%%%%%%%%%%%%%%%%%
 %%%%%%%%%%%%%%%%%%%%%%%%%%%%
 %%%%%%%%%%%%%%%%%%%%%%%%%%%%
 %%%%%%%%%%%%%%%%%%%%%%%%%%%%

\section{Collider and cosmological constraints \label{sec:pheno}} 
 
 The scenarios studied in this paper  involve additional scalar singlets with a large multiplicity and/or strong couplings with the Higgs, so that despite their lack of interactions with the SM gauge fields they could potentially yield measurable effects in colliders, especially linear ones,  as they would able to probe the Higgs couplings with high precision. In the following, possible collider and cosmological constraints are summarized. 
 
 The physical masses of the singlet fields are
 \begin{align*}
  \tilde m^2=m^2_S+\frac{\lambda_{SH}}{2}v^2.
 \end{align*}
 Even in the $m^2_S=0$ case and with a negligible barrier between the origin and the electroweak vacuum at zero temperature, as in the Coleman-Weinberg scenarios, the results of the previous section show that even for as many as twelve singlets the coupling $\lambda_{SH}$  is greater than one. Thus, the singlet fields are heavier than the Higgs, so that there are no constraints coming from its invisible width. Additional collider constraints may result from jets plus missing-energy signals, coming from vector-boson-fusion production of off-shell Higgses decaying into singlets. The following table presents results for the 13 TeV LHC cross-sections for these processes in terms of the number of singlets and the coupling $\lambda_{SH}$, which were calculated using MadGraph 5 \cite{Alwall:2011uj}:
 
 \begin{minipage}\linewidth
 \begin{center}
\begin{tabular}{c|c|c|c|c}
N & 1 & 3 & 6 & 12\\
\hline
$\lambda$ & 6.89 & 4.10 & 2.93 & 2.10\\
\hline
$\sigma$ (fb)& 0.33 & 1.34  & 3.04 & 6.77\\
\end{tabular}
\captionof{table}{LHC cross sections for processes yielding two-jets plus two-singlets.}\label{table:xs}
\end{center}
\end{minipage}

In order to compare with current limits from jets plus missing-energy searches, the cuts of the 2-jet searches in ref.~\cite{TheATLAScollaboration:2013fha} were implemented on top of simulated events, obtained by running Pythia 6 \cite{Sjostrand:2001yu} and PGS \cite{PGS} in conjunction with MadGraph. These cuts are: $E^{\rm miss}_{\rm T}>160\, {\rm GeV}, \,p_T(j_1)>130\, {\rm GeV},\,  p_T(j_2)>60\, {\rm GeV}, \,\Delta\Phi(j_i,E^{\rm miss}_{\rm T})>0.4$, where $E^{\rm miss}_{\rm T}$ is the missing transverse energy,  $p_T(j_i)$ the magnitude of the transverse  momentum of jet $i$, and $\Delta\Phi(j_i,E^{\rm miss}_{\rm T})$ the smallest azimuthal separation between the missing energy and the jet momenta. The search region labeled as ``A-loose'' in ref.~\cite{TheATLAScollaboration:2013fha}  further demands $E_{\rm T}^{\rm miss}/m_{\rm eff} (2 j)>0.2, \,m_{\rm eff}(2 j)>1000 \,{\rm GeV}$, where $m_{\rm eff}(2 j)$ denotes the effective mass of the 2-jet system. Finally, the ``A-medium'' search region requires $E_{\rm T}^{\rm miss}/\sqrt{H_T}> 15 \,{\rm GeV}^{1/2}$, where $H_T$ represents the scalar sum of all transverse jet momenta, and  $m_{\rm eff}(2 j)>1600 \,{\rm GeV}$. The implementation of these cuts reduces the  cross-sections in table~\ref{table:xs} by a factor of order 100(1000) for the loose (medium) search regions, rendering them several hundred times beyond the reported cross-section limits at the 95\% confidence level of 66.07(2.52)fb.

\begin{figure}[h!]\centering
  \includegraphics[scale=.67]{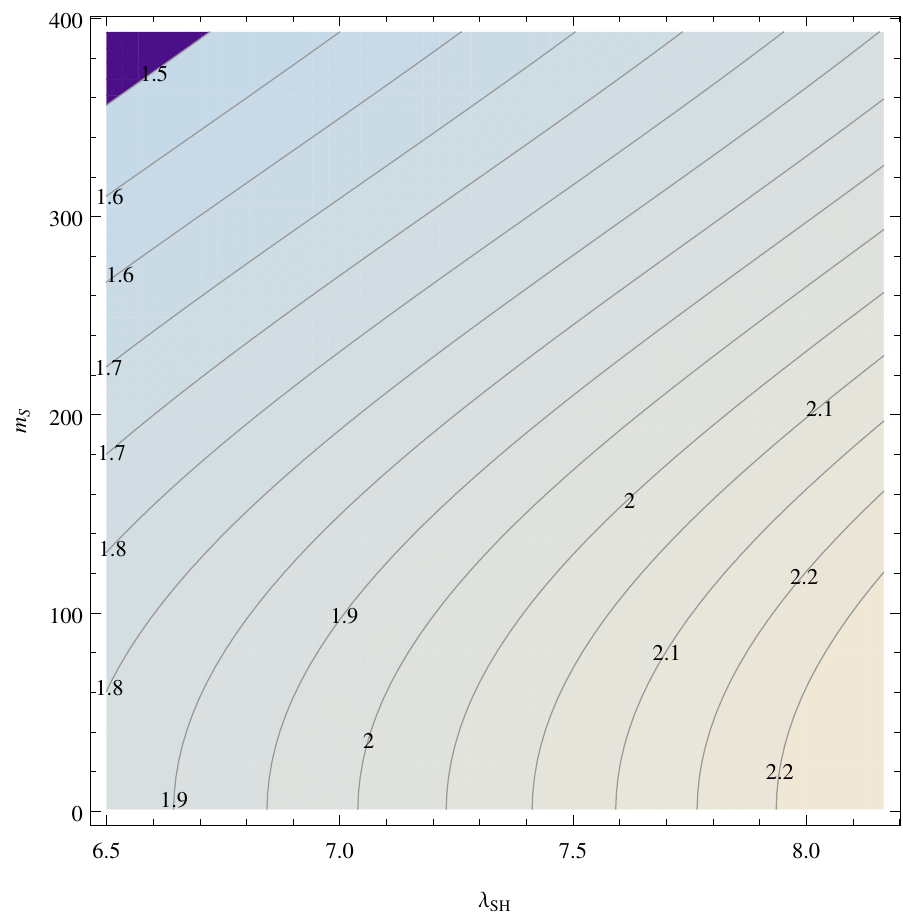}
     \includegraphics[scale=.67]{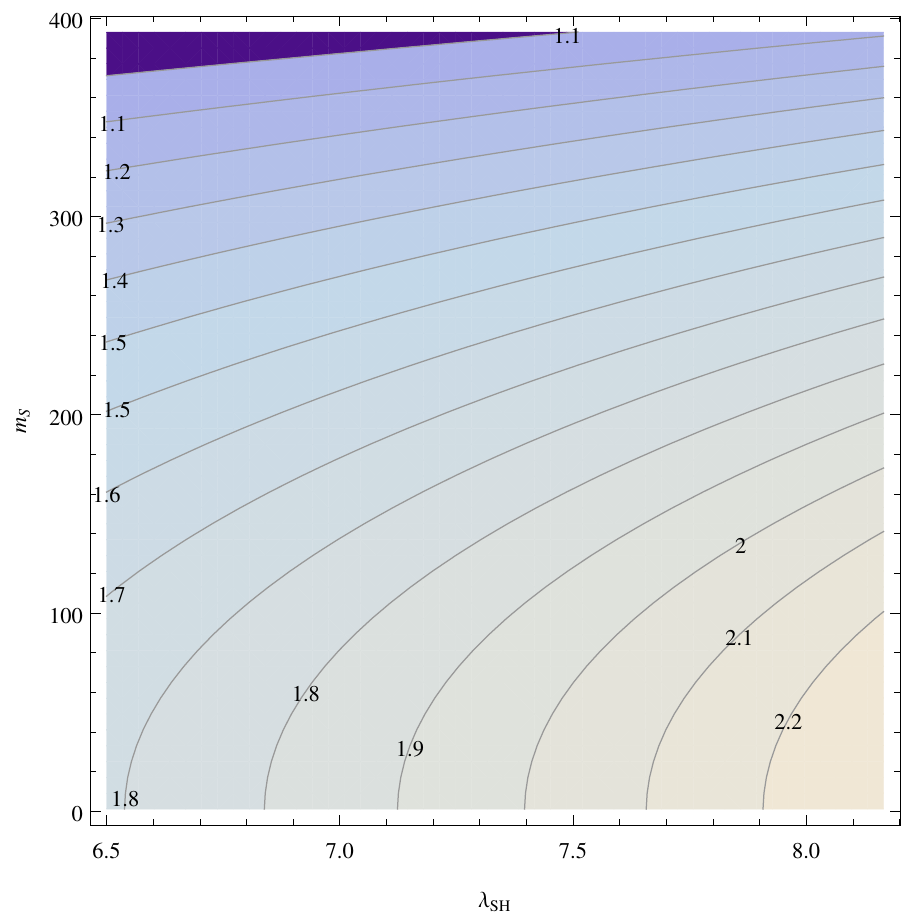}
      \includegraphics[scale=.67]{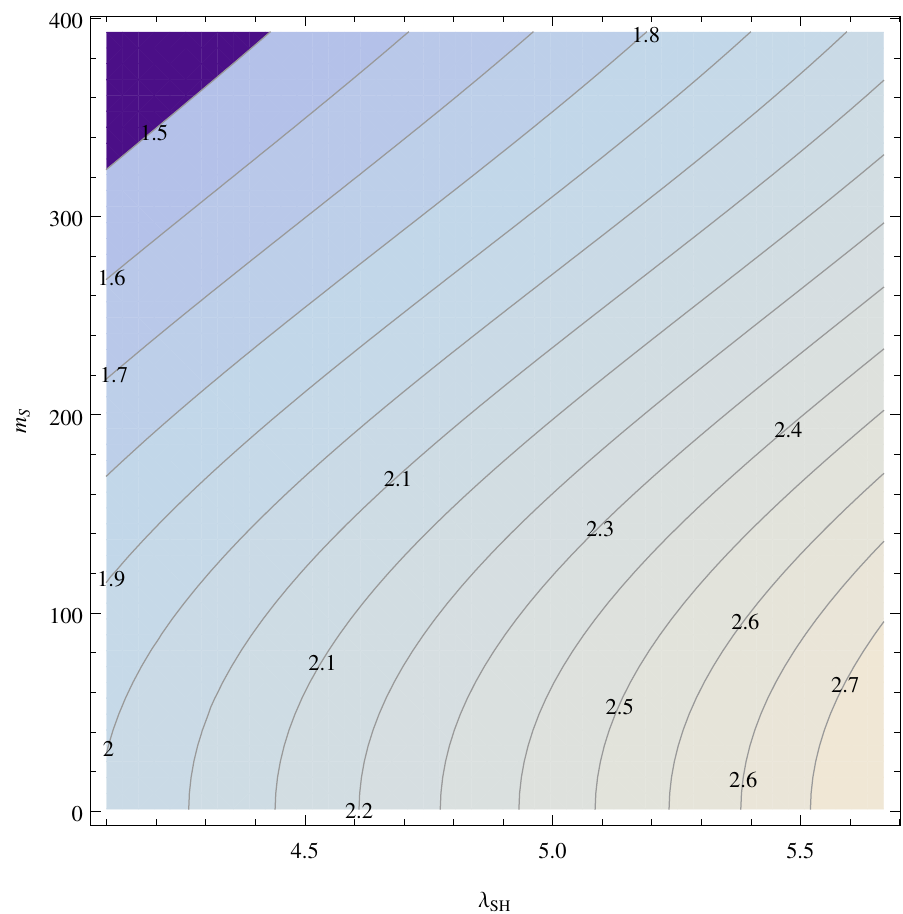}
    \includegraphics[scale=.67]{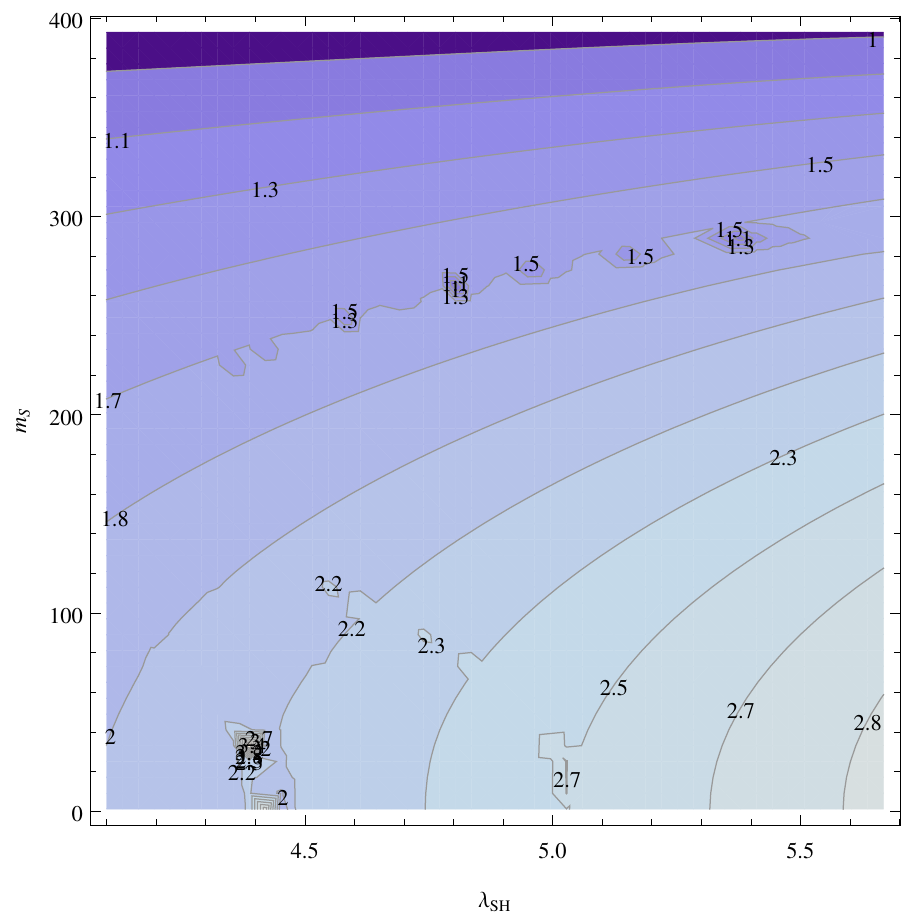}
    \includegraphics[scale=.67]{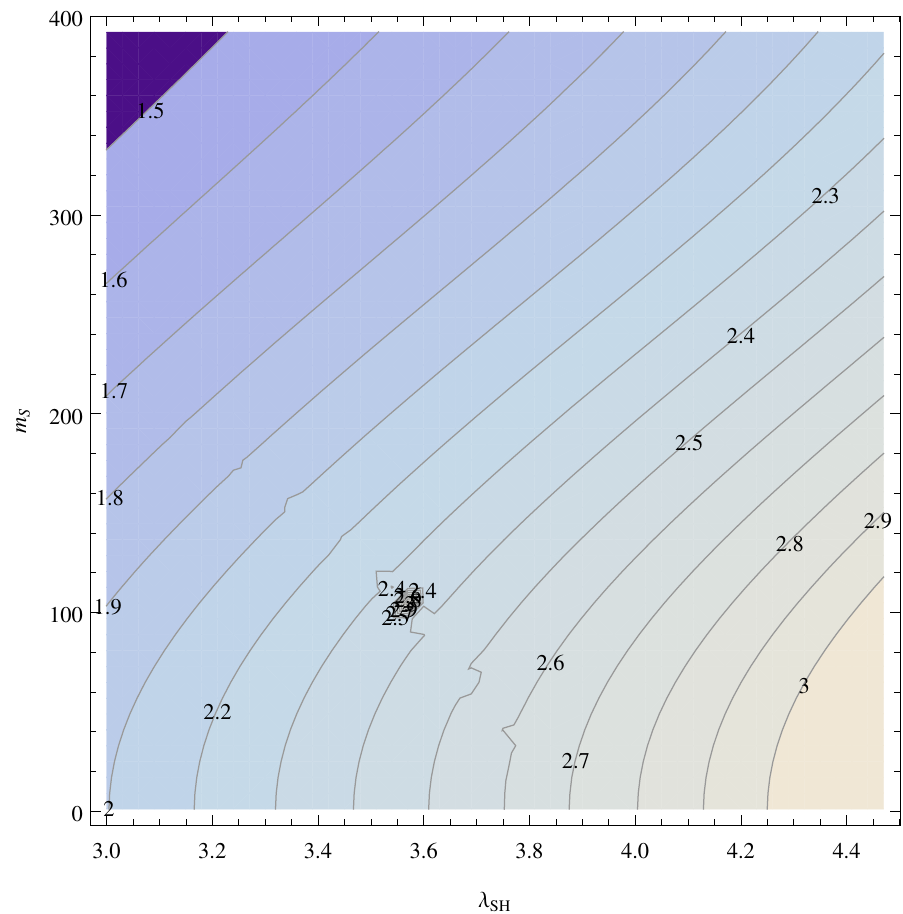}
    \includegraphics[scale=.67]{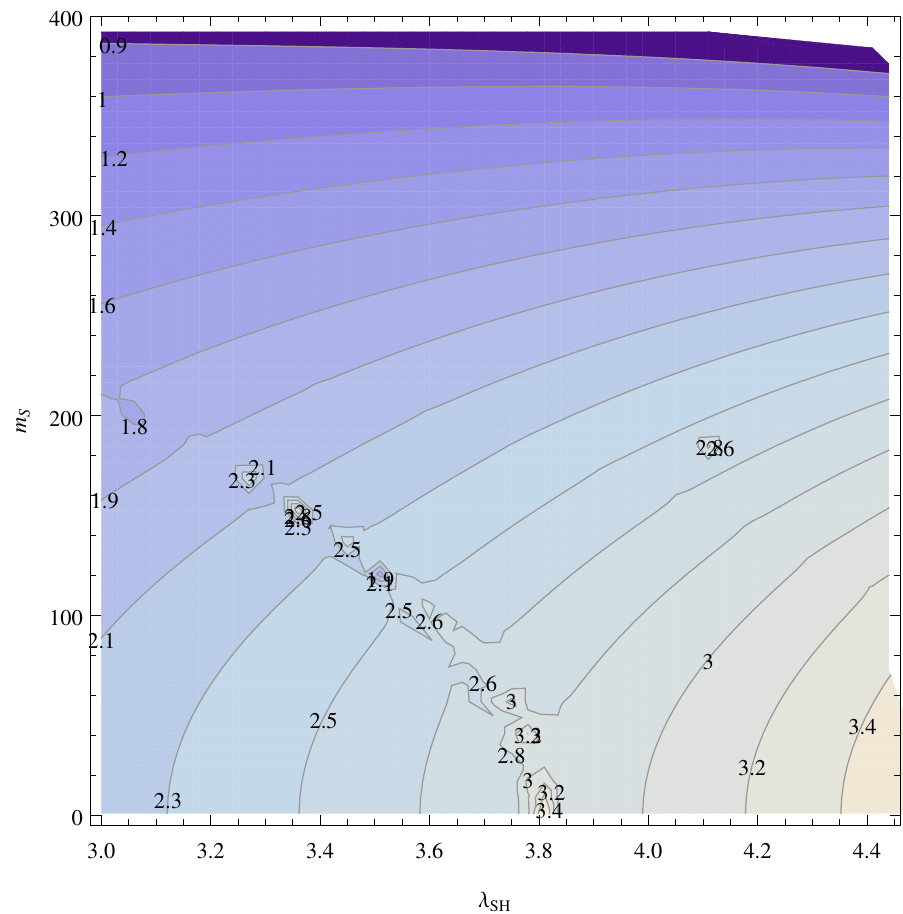}
\caption{\label{fig:ah}Enhancement of the cubic Higgs coupling with respect to its Standard Model value ($a_h/a^{SM}_h$), in models  $N=1$ (top), $N=3$ (middle row) and $N=6$ (bottom), at one-loop (left), and two-loops (right).  $\lambda_S$ was fixed at 0.1.}
\end{figure}

Aside from missing-energy signals, another possible collider signature would be a deviation from the Standard Model in the Higgs cubic coupling  \cite{Noble:2007kk}.  The scenarios studied here typically have  $\lambda\lesssim0$ and  $m^2_H\gtrsim0$ at low scales, in contrast with the SM, so that anomalous Higgs cubic couplings are to be expected. Neglecting the contributions to the effective potential coming from the Higgs and fermions other than the top,  and substituting $m^2_H$ in terms of the rest of parameters by making use of condition \eqref{eq:m2H},  one obtains the following approximate expression for  the one-loop value of the cubic coupling at $\mu=v$:

 \begin{align*}a_h\equiv&\frac{d^3 V(h)}{dh^3}=3\lambda \tilde v+\frac{1}{16\pi^2}\left\{\frac{81}{200} g_1^4 \tilde{v} \log \left(\frac{m^2_Z(\tilde v)}{v^2}\right)+\frac{27}{20} g_2^2 g_1^2 \tilde{v} \log \left(\frac{m^2_Z(\tilde v)}{v^2}\right)\right.\\
 %%%%%
 &+\frac{9}{4} g_2^4 \tilde{v} \log \left(\frac{m^2_W(\tilde v)}{v^2}\right)+\frac{9}{8} g_2^4 \tilde{v} \log \left(\frac{m^2_Z(\tilde v)}{v^2}\right)+\frac{27}{50} g_1^4 \tilde{v}+\frac{9}{5} g_2^2 g_1^2 \tilde{v}+\frac{9}{2} g_2^4 \tilde{v}\\
 %%%%
 &+\frac{27}{2} \lambda ^2 \tilde{v} \log \left(\frac{m^2_h(\tilde v)}{v^2}\right)+\frac{27 \lambda ^3 \tilde{v}^3}{2m^2_H(\tilde v)}+3N \tilde{v} \lambda _{\text{SH}}^2 \log \left(\frac{m^2_S(\tilde v)}{v^2}\right)+\frac{2N\tilde{v}^3 \lambda _{\text{SH}}^3}{2 m^2_S(\tilde v)}\\
 %%%%
 &\left.-18 \tilde{v} y_t^4 \log \left(\frac{\tilde{v}^2 y_t^2}{2 v^2}\right)-12 \tilde{v} y_t^4-\frac{27 \lambda ^2 \tilde{v}}{4}\right\}.
 \end{align*}
As in \S~\ref{sec:VCW}, all couplings are understood to be evaluated at the scale $\mu=v(m_Z)$,  $\tilde v$ is defined  in eq.~\eqref{eq:lambdam2H}, and the field-dependent masses are given in eq.~\eqref{eq:m2H}. For large values of $\lambda_{SH}$, the positive contributions proportional to $\lambda^3_{SH}$ will dominate, providing a significant enhancement with respect to the Standard Model value of $a_h^{SM}=167$ GeV (obtained from the SM two-loop, RG-improved effective potential). Fig.~\ref{fig:ah} shows the enhancement of the cubic coupling with respect to its SM value for different values of $N$,  at one and two-loops. In both cases, the field-dependent masses of the Goldstone modes, whose effect is subleading, were set to zero in order to avoid the appearance of discontinuities in $a_h$ signaling the need to perform a  proper resummation. Enhancements of around 100\% are typical for low values of $m^2_S$, making measurements of the Higgs cubic interaction a powerful probe of these scenarios. The deviations with respect to the SM value are larger than the expected precision of 20\% achievable at prospective linear colliders \cite{Djouadi:2007ik}; also, cubic couplings enhanced by a factor of 2 could be measurable at the LHC \cite{Dolan:2012rv,Baglio:2012np}. The results obtained here are larger than the  corresponding $N=6$ values in ref.~\cite{Espinosa:2008kw} for scenarios with a $125$ GeV Higgs mass and a nonzero potential barrier. In this work, the minimum relative enhancement of the cubic coupling with respect to the SM in the $m^2_S=0$ case is estimated to be be of the order of 75\%, as opposed to the value around 100\% found here. The kinks in some of the plots in fig.~\ref{fig:ah} are caused by singularities in the one-loop logarithms or two-loop functions when some field-dependent masses become zero. This could be avoided by properly resumming Goldstone effects, as in refs.~\cite{Martin:2014bca} and \cite{Elias-Miro:2014pca}; such a treatment lies beyond the scope of this work, and the results should not be affected outside the regions of singularity.

An additional signature of these scenarios is a modification  of Higgs associated-production cross sections due to wavefunction renormalization effects. As noted in refs.~\cite{Craig:2013xia,Englert:2013tya,Katz:2014bha}, this is a typical effect in new physics scenarios involving modified Higgs couplings --particularly those giving rise to strong first-order electroweak phase transitions-- even when the new fields are singlets. Following ref.~\cite{Craig:2013xia}, the fractional deviation with respect to the SM prediction in the Higgs associated-production cross-section in an electron-positron collider  is given by
\begin{align*}
 \delta \sigma_{Zh}=&\frac{N\lambda_{SH}^2v^2}{16\pi^2m^2_h}\left(1+F\left[\frac{m^2_h}{4\tilde m^2_S}\right]\right),\\
 %%%%%
 F[x]=&\frac{1}{4\sqrt{x(x-1)}}\log\left(\frac{1-2x-2\sqrt{x(x-1)}}{1-2x+2\sqrt{x(x-1)}}\right).
\end{align*}
The following table presents values for $\delta \sigma_{Zh}$ in the limiting Coleman-Weinberg scenarios:

 \begin{minipage}\linewidth
 \begin{center}
\begin{tabular}{c|c|c|c|c}
N & 1 & 3 & 6 & 12\\
\hline
$\lambda_{SH}$ & 6.55 & 4.19 & 3.18 & 2.43\\
\hline
$\delta\sigma_{ZH}$ & -0.014 & -0.027  & -0.042 & -0.064\\
\end{tabular}
\captionof{table}{Fractional deviations in Higgs associated-production cross-sections.}\label{table:Zhxs}
\end{center}
\end{minipage}

For lower values of $N$, the mass of the singlet fields increases and their effect decouples. Raising $\lambda_{SH}$ for fixed $N$ enhances the fractional deviation, so that the values in the table correspond to minimal deviations in scenarios with nonzero barriers. The deviations are greater than the projected $O(0.5\%)$ precision in cross-section measurements in  prospective linear colliders \cite{Peskin:2012we,Klute:2013cx}.

Moving on to cosmological considerations, the reader should be reminded that the potential of eq.~\eqref{eq:Vscalar} was obtained under the simplifying assumption of a global symmetry. This  enforce the stability of the singlets, which could become dark-matter candidates; however, their strong Higgs-portal couplings make them annihilate too efficiently to generate the correct relic abundance \cite{Espinosa:2008kw}. The tree-level, non-relativistic cross-section for the annihilation of one of the particles in the multiplet $S$ into Higgs, $W$ and $Z$ bosons and fermions is given by
\begin{align}
\nonumber
 \sigma^{ss\rightarrow hh}_{NR}(E_{CM})v=&\frac{\lambda_{SH}^2}{64\pi \tilde m_{S}^3}\sqrt{\tilde m_{S}^2-m_h^2}\left(1+\frac{9\lambda^2v^4}{(4\tilde m^2_S-m^2_h)^2}+\frac{4\lambda_{SH}^2 v^4}{m_h^4}+\frac{6\lambda v^2}{4\tilde m^2_S-m^2_h}\right.\\
 %%%%%
 \nonumber&\left.-\frac{12\lambda_{SH}\lambda v^4}{(4\tilde m^2_S-m^2_h)m^2_h}-\frac{4\lambda_{SH}v^2}{m^2_h}\right)+\left\{\frac{\lambda_{SH}^2 m^4_W}{8\pi\tilde m^3_S(4\tilde m^2_S-m^2_h)^2}\sqrt{\tilde m_{S}^2-m_W^2}\left(2\right.\right.\\
 %%%%%
 \label{eq:sigmaan}&\left.\left.+\frac{(2\tilde m^2_S\!-\!m^2_Z)^2}{m^4_W}\right)+\frac{1}{2}(m_W\leftrightarrow m_Z)\right\}+\sum_f \frac{n_f\lambda_{SH}^2 m^2_f}{4\pi \tilde m^3_S(4\tilde m^2_S-m^2_h)^2}(\tilde m_{S}^2-m_f^2)^{\frac{3}{2}},
\end{align}
where $E_{CM}$ is the center-of-mass energy. The largest relic abundances will be obtained for a large number of singlets with small annihilation cross-sections. In the scenarios studied here both things go together as larger $N$ requires smaller values of $\lambda_{SH}$ to generate a barrier. In the case with smallest possible $\lambda_{SH}$ considered in this paper, as in the $N=12$ Coleman-Weinberg scenario with $\lambda_{SH}\sim2.4,\lambda\sim0.3$ and involving 24 real singlets with $m_{S,phys}\sim v$, this gives $\sigma^{ss\rightarrow hh}_{NR}(2m_{S,phys}) v\sim1.2\cdot10^{-21}\frac{cm^3}{s}$. Approximating the thermally-averaged  annihilation cross-section by eq.~\eqref{eq:sigmaan} and using it as an input in the usual analytical approximations for the freezout temperature and relic abundance resulting from s-channel annihilations (see for example ref.~\cite{Kolb:1990vq}), then it is easily seen that the singlets decouple at a temperature $T_f\sim\frac{v}{33}$ and give rise to a total relic abundance of $\Omega_{DM}(N=12)\sim1.5\cdot10^{-4}$, well below the observed dark-matter abundance. The fact that $\lambda^{CW}_{SH}(N)\sim \frac{1}{\sqrt{N}}$ decreases slowly for large values of $N$ implies that $\Omega\sim0.2$ can not be reached even for $N\sim 100$. It should be noted that allowing for different sectors of scalars with different strength of their Higgs-portal couplings allows to achieve both electroweak-symmetry breaking \`{a} la Coleman-Weinberg and dark-matter candidates with the correct relic abundance; see for example ref.~\cite{Guo:2014bha}.

To end this section, it is worth pointing out that  new CP violating phases are allowed if the assumption of a global symmetry is abandoned. This opens the possibility that these models do not only provide a strong first-order phase transition, but also enough CP violation to explain the baryon asymmetry.  Nevertheless, for these phases to have any  effect, nonzero singlet VEVs would be typically required, giving rise to changing CP phases along a bubble configuration that cannot be rotated away. This lies beyond the scope of this paper, in which singlet VEVs were ignored and the global symmetry was enforced.

 %%%%%%%%%%%%%%%%%%%%%%%%%%%%
 %%%%%%%%%%%%%%%%%%%%%%%%%%%%
 %%%%%%%%%%%%%%%%%%%%%%%%%%%%
 %%%%%%%%%%%%%%%%%%%%%%%%%%%%

  \section{Conclusions}

  This paper presents a detailed analysis of models of radiative electroweak symmetry breaking in which scalar fields with strong Higgs-portal couplings give rise to Higgs vacua separated from the origin by a barrier at zero temperature \cite{Espinosa:2007qk}. These scenarios have many properties of interest. They cure the instability of the Standard Model potential, they ensure strong first-order electroweak phase transitions --a necessary ingredient for viable models of electroweak baryogenesis-- and in the limit of zero barrier height they  provide realizations of electroweak symmetry breaking \`{a} la Coleman-Weinberg, thus providing a possible dynamical explanation for the origin of the electroweak scale.
  
  For these scenarios to be viable, either large couplings or a large number of singlets are needed.  Previous analysis in the literature  \cite{Espinosa:2007qk,Espinosa:2008kw} focused on stop-inspired scenarios with 12 real singlets, which raised the question of whether the large value involved for $\lambda_{SH} N\sim O(10)$ (where $\lambda_{SH}$ is the Higgs-portal coupling and $N$ the number of complex singlets) allowed for trustworthy perturbative calculations. Moreover, it was unclear whether scenarios with fewer singlets could also exhibit similar properties while remaining calculable. In this paper, these issues were tackled by considering theories with an arbitrary number of complex scalars and calculating the renormalization-group improved effective potential at two-loop order, reproducing the correct Higgs VEV and mass. Finite temperature effects were also considered, including the dominant two-loop corrections involving the Higgs-portal coupling.
  
  The results show that two-loop corrections are under control around the electroweak scale for all values of $N$, as follows from examining the behavior of the effective potential under changes of scale. The value of the Higgs-portal couplings needed to produce a certain potential barrier go as $\lambda_{SH}\sim\frac{1}{\sqrt{N}}$, (with $\lambda\gtrsim6.5$ for $N=1$ under the requirement of a nonzero barrier height), while perturbative corrections involving the Higgs-portal coupling are expected to depend on $\lambda_{SH} N\sim\sqrt{N}$. This suggests that scenarios with smaller $N$, although requiring larger couplings, are under better perturbative control, which is confirmed by the numerical results. Thus, scenarios with small $N$ are not only viable, but also better behaved.
  
  The calculations at finite temperature show that these models have strong first-order electroweak phase transitions even in the limiting Coleman-Weinberg scenarios. Collider and cosmological constraints were also analyzed; regarding the former, it was shown the jets plus missing-energy signals coming from vector-boson-fusion production of off-shell Higgses decaying into singlets are significantly below the current limits, and yet the cubic Higgs coupling and the Higgs associated production cross section depart significantly from their Standard Model values and could be probed at future linear colliders. The Higgs cubic coupling is typically enhanced by 100\% --more than the previous estimate in ref.~\cite{Espinosa:2008kw}-- while the Higgs associated-production cross-section exhibits at least a 1.4\% suppression. These effects are  greater than the expected precision of the corresponding measurements in linear colliders, which are of the order of 20\% and 0.5\%, respectively \cite{Djouadi:2007ik,Peskin:2012we,Klute:2013cx}. As pertains to cosmological constraints, in models with stable singlets the strong Higgs-portal coupling gives rise to annihilation cross-section that are too large to reproduce the observed dark matter relic abundance.
  
  Finally, although the large Higgs-portal couplings of the scenarios studied here make them nonperturbative near the TeV scale, it is possible to modify them so as to preserve perturbativity up to the Planck scale by enriching the interactions in the hidden sector, as in ref.~\cite{Dermisek:2013pta}.

\section*{Acknowledgements}I am indebted to Gordan Krnjaic, David Morrisey, Philip Schuster, Brian Shuve and Itay Yavin for useful conversations. I acknowledge support from the Spanish Government through grant FPA2011-24568. Research at the Perimeter Institute is supported in part by the Government of Canada through Industry Canada, and by the Province of Ontario through the Ministry of Research and Information (MRI).
 %%%%%%%%%%%%%%%%%%%%%%%%%%%%
 %%%%%%%%%%%%%%%%%%%%%%%%%%%%
 %%%%%%%%%%%%%%%%%%%%%%%%%%%%
 %%%%%%%%%%%%%%%%%%%%%%%%%%%%
\appendix

\section{Two-loop beta functions and Higgs anomalous dimension\label{app:RG}}
 %%%%%%%%%%%%%%%%%%%%%%%%%%%%
 %%%%%%%%%%%%%%%%%%%%%%%%%%%%
 %%%%%%%%%%%%%%%%%%%%%%%%%%%%
 %%%%%%%%%%%%%%%%%%%%%%%%%%%%
 This appendix displays the 2-loop beta functions and the Higgs anomalous dimension for the theory obtained by adding a complex multiplet of $N$ singlet scalars to the SM, and considering the scalar interactions of eq.~\eqref{eq:Vscalar}. The formulae were obtained by applying the results of refs.~\cite{Machacek:1983tz,Machacek:1983fi,Machacek:1984zw,Luo:2002ti}. For interactions already present in the Standard Models, the beta functions are given in terms of their value in the Standard Model --denoted with a superscript ``SM'', see refs.~\cite{Machacek:1983tz,Luo:2002ey}. In the expressions that follow complex phases were omitted, as well as all Yukawa couplings except for the third-generation diagonal ones, denoted by $y_t,y_b,y_\tau$. The hypercharge, weak and strong couplings are denoted by $g_1,g_2,g_3$, and $g_1$ is assumed to be in the GUT normalization.
  \begin{align*}
   \beta_{g_1}=&\beta_{g_1}^{SM},\\
   %%%%%
   \beta_{g_2}=&\beta_{g_2}^{SM},\\
   %%%%%
   \beta_{g_3}=&\beta_{g_3}^{SM},\\
   %%%%%
   \beta_{y_t}=&\beta_{y_t}^{SM}+\frac{N\lambda _{SH}^2 y_t}{512\pi^4},\\
   %%%%%
   \beta_{y_b}=&\beta_{y_b}^{SM}+\frac{N\lambda _{SH}^2y_b}{512\pi^4},\\
    %%%%%
   \beta_{y_\tau}=&\beta_{y_\tau}^{SM}+\frac{N\lambda _{SH}^2 y_\tau}{512\pi^4},\\
    %%%%%
   \beta_\lambda=&\beta_\lambda^{SM}+\frac{N\lambda _{SH}^2}{8\pi^2}+\frac{N}{(16\pi^2)^2}\left[ -10\lambda  \lambda _{SH}^2-8 \lambda _{SH}^3\right],\\
    %%%%%
   \beta_{m^2_H}=&\beta_{m^2_H}^{SM}+\frac{N}{8\pi^2}m_S^2 \lambda _{SH}+\frac{N}{(16\pi^2)^2}\left[-m_H^2 \lambda _{SH}^2-4m_S^2  \lambda _{SH}^2\right],\\
   %%%%%%
   %%%%%%
   \gamma_H=&\gamma_H^{SM}+\frac{N}{512\pi^4}\lambda _{SH}^2,\\
   %%%%%%
   %%%%%%
   %%%%%%
   \beta_{\lambda_S}=&\frac{1}{16\pi^2}\left[(8+2N) \lambda _S^2+4 \lambda _{SH}^2\right]
+\frac{1}{(16\pi^2)^2}\left[\lambda_{SH}^2 \left(-24 y_b^2+\frac{24 g_1^2}{5}+24 g_2^2-24 y_t^2-8 y_{\tau }^2\right)\right.\\
%%%%%
&\left.-(42+18N) \lambda_S^3-20 \lambda_S \lambda_{SH}^2-16 \lambda_{SH}^3\right],\\
%%%%
%%%%
%%%%
%%%%
%%%%
%%%%
\beta_{\lambda_{SH}}=&\frac{1}{16\pi^2}\left[\lambda _{SH} \left(6 y_b^2-\frac{1}{10} 9 g_1^2-\frac{9 g_2^2}{2}+6 \lambda +(2+2N) \lambda _S+6 y_t^2+2 y_{\tau }^2\right)+4 \lambda _{SH}^2\right]\\
%%%%
&+\frac{1}{(16\pi^2)^2}\left[\lambda _{SH} \left(\lambda  \left(\frac{36 g_1^2}{5}+36 g_2^2-36 y_t^2-36 y_b^2-12 y_{\tau }^2\right)+g_1^2 \left(\frac{5 y_b^2}{4}+\frac{9 g_2^2}{8}+\frac{17 y_t^2}{4}\right.\right.\right.\\
%%%%
&\left.+\frac{15 y_{\tau }^2}{4}\right)+g_2^2 \left(\frac{45 y_b^2}{4}+\frac{45 y_t^2}{4}+\frac{15 y_{\tau }^2}{4}\right)+g_3^2 \left(40 y_b^2+40 y_t^2\right)-21 y_b^2 y_t^2-\frac{27 y_b^4}{2}+\frac{1671 g_1^4}{400}\\
%%%%
&\left.-\frac{145 g_2^4}{16}-15 \lambda ^2-(5+5N) \lambda _S^2-\frac{27 y_t^4}{2}-\frac{9 y_{\tau }^4}{2}\right)+\lambda _{SH}^2 \left(-12 y_b^2+\frac{3 g_1^2}{5}+3 g_2^2-36 \lambda\right.\\
%%%
&\left.\left. -12(1+N)\lambda _S-12(1+N)\lambda _S-12 y_t^2-4 y_{\tau }^2\right)-(10+N) \lambda _{SH}^3\right],\end{align*}\begin{align*}
%%%%%
%%%%%
%%%%%
\beta_{m^2_S}=&\frac{1}{16\pi^2}\left[4 m_H^2 \lambda _{SH}+2(1+N)m^2_S\lambda _S\right]+\frac{1}{(16\pi^2)^2}\left[m_H^2 \left(\lambda _{SH} \left(-24 y_b^2+\frac{24 g_1^2}{5}+24 g_2^2\right.\right.\right.\\
%%%
&\left.\left.\left.-24 y_t^2-8 y_{\tau }^2\right)-8 \lambda _{SH}^2\right)+m^2_S\left(-5(1+N) \lambda _S^2-2 \lambda _{SH}^2\right)\right].
  \end{align*}

 %%%%%%%%%%%%%%%%%%%%%%%%%%%%%%%%%%%%%%%%%%%%%%%%%%%%%%%%%%%%%%%%%%%%%%%%%%%%%%%%%%%%%%%%%%%%%%%%%%%%%%%%%
 %%%%%%%%%%%%%%%%%%%%%%%%%%%%%%%%%%%%%%%%%%%%%%%%%%%%%%%%%%%%%%%%%%%%%%%%%%%%%%%%%%%%%%%%%%%%%%%%%%%%%%%%%
 %%%%%%%%%%%%%%%%%%%%%%%%%%%%%%%%%%%%%%%%%%%%%%%%%%%%%%%%%%%%%%%%%%%%%%%%%%%%%%%%%%%%%%%%%%%%%%%%%%%%%%%%%
 %%%%%%%%%%%%%%%%%%%%%%%%%%%%%%%%%%%%%%%%%%%%%%%%%%%%%%%%%%%%%%%%%%%%%%%%%%%%%%%%%%%%%%%%%%%%%%%%%%%%%%%%%
 %%%%%%%%%%%%%%%%%%%%%%%%%%%%%%%%%%%%%%%%%%%%%%%%%%%%%%%%%%%%%%%%%%%%%%%%%%%%%%%%%%%%%%%%%%%%%%%%%%%%%%%%%
 %%%%%%%%%%%%%%%%%%%%%%%%%%%%%%%%%%%%%%%%%%%%%%%%%%%%%%%%%%%%%%%%%%%%%%%%%%%%%%%%%%%%%%%%%%%%%%%%%%%%%%%%%

\section{Two-loop finite-temperature scalar diagrams\label{app:diagrams}}
 %%%%%%%%%%%%%%%%%%%%%%%%%%%%
This appendix provides formulae for the dominant two-loop finite-temperature contributions to the effective potential, which are  driven by  the Higgs-portal coupling $\lambda_{SH}$. Calculations were done in dimensional regularization with $D=4-2\epsilon$ in the $\overline{\text{MS}}$ scheme. The corresponding diagrams are given in fig.~\ref{fig:scalardiagrams}, where double lines correspond to singlets and single ones to fields in the Higgs multiplet. In order to isolate the finite temperature contribution, calculations were done in the imaginary time formalism,  making repeated use of the identity

\begin{align}
\label{eq:p0}
&T\sum_{n=-\infty}^\infty f(p_0=i\omega_n)=\\
\nonumber&\frac{1}{2\pi i}\int_{-i\infty}^{i\infty}dp_0\frac{1}{2}[f(p_0)+f(-p_0)]+\frac{1}{2\pi i}\int_{-i\infty+\epsilon}^{i\infty+\epsilon}dp_0[f(p_0)+f(-p_0)]\frac{1}{e^{\beta p_0}-1},\,\,\omega_n=\frac{2\pi n}{\beta}.
\end{align}
The results can be expressed in terms of the following one and two-loop dimensionally regularized integrals:
\begin{align*}
 I[m^2]=\,&T\sum_{n=-\infty}^\infty\int\frac{d^{D-1}p}{(2\pi)^{D-1}}\frac{1}{\omega_n^2+p^2+m^2},\\
 %%%%
 H[m^2,M^2]=\,&T^2\sum_{n,m=-\infty}^\infty\int\frac{d^{D-1}p}{(2\pi)^{D-1}}\frac{d^{D-1}q}{(2\pi)^{D-1}}\frac{1}{\omega_n^2+p^2+m^2}\frac{1}{(\omega_m+\omega_n)^2+(p+q)^2+m^2}\\
 %%%%
 &\times\frac{1}{\omega_m^2+q^2+M^2}\left(\frac{e^{\gamma_E}\mu}{4\pi}\right)^{2\epsilon},
\end{align*}
where $D=4-2\epsilon$.

Some relevant formulae for the integrals are given next. They have been obtained by successively applying eq.~\eqref{eq:p0}  and by performing the complex integrals by suitable closures of contours.
The zero-temperature contributions and the temperature-dependent corrections are denoted by suffixes 0 and $T$, respectively. Superindices are used to denote the different orders in a Laurent expansion in $\epsilon$.

\begin{align}
 \nonumber I[m^2]=&I_0[m^2]+I_T[m^2];\\
 %%%%
 \nonumber I_0[m^2]=&\frac{1}{\epsilon}I^{-1}_0[m^2]+I^0_0[m^2]+O(\epsilon),\,\,I_T[m^2]=I^0_T[m^2]+\epsilon I_T^1[m^2]+O(\epsilon^2),\\
 %%%%%%%%%%
 \nonumber I^{-1}_0[m^2]=&-\frac{m^2}{16\pi^2},\quad I_0^0[m^2]=-\frac{m^2}{16\pi^2}\left(1-\gamma_E-\log\frac{m^2}{4\pi}\right),\\
 %%%%%%%%%%
 \nonumber I^0_T[m^2]=&\frac{1}{2\pi^2}\int_0^\infty dp\frac{p^2}{\tomegap}n(\tomegap),\\ 
 %%%%%%%%%
 \nonumber I_T^1[m^2]=&\frac{1}{2\pi^2}\int_0^\infty dp\frac{p^2}{\tomegap}n(\tomegap)\left(2-\gamma_E-\log \frac{p^2}{\pi}\right),\\
 %%%%%%%
 \label{eq:ints}H_T[m^2,M^2]=&H^0_{T,1}[m^2,M^2]+H^0_{T,2}[m^2,M^2]+O(\epsilon),\\
 %%%%%%%%%%
 \nonumber H^0_{T,1}[m^2,M^2]=&-\frac{1}{16\pi^4}\int_0^1 dx\int_0^\infty dp\frac{p^2n(\tomegap)}{\tomegap}(\log p^2(M^2(1-x)+m^2 x^2))\\
 %%%%%
 \nonumber &+\frac{1}{8\pi^4}\left(1+\log\frac{\mu}{2}\right)\int_0^\infty dp\frac{p^2n(\tomegap)}{\tomegap},\\
 %%%%%%%%
 %%%%%%%%
\nonumber  H^0_{T,2}[m^2,M^2]=&-\frac{1}{16\pi^4}\int_0^\pi\!\! d\theta\sin\theta\int_0^\infty\!\! dp dq\,p^2q^2\Big\{\frac{1}{\tomegap\tomegaq}\Big[\frac{n(\tomegap)n(\tomegaq)}{(\tomegap\!+\tomegaq)^2-\tomegapq^2}+\frac{n(\tomegap)n(\tomegaq)}{(\tomegap\!-\tomegaq)^2-\tomegapq^2}\Big]\\
 %%%
\nonumber  &+\frac{n(\tomegap+\tomegapq)}{\tomegap\tomegapq((\tomegap+\tomegapq)^2-\tomegaq^2)}\Big[n(\tomegap)+n(-\tomegap)\Big]+\frac{n(\tomegaq)}{\tomegaq\tomegapq}\Big[\frac{n(\tomegapq-\tomegaq)}{(\tomegaq-\tomegapq)^2-\tomegap^2}\\
 %%%%%
 \nonumber &+\frac{n(\tomegapq+\tomegaq)}{(\tomegaq+\tomegapq)^2-\tomegap^2}\Big]\Big\},
\end{align}
where the following shorthands were used,
\begin{align*}
 &\tomegap\equiv\sqrt{p^2+m^2},\\
 %%%%
  &\tomegaq\equiv\sqrt{q^2+m^2},\\
 %%%%.
 & \tomegapq\equiv\sqrt{p^2+q^2+2 p q \cos\theta+M^2},\\
 %%%%%
 &n(x)\equiv\frac{1}{e^{\beta x}-1}.
\end{align*}
The counterterm diagrams of fig.~\ref{fig:scalardiagrams} yield the following $\lambda_{SH}$- and temperature-dependent contributions:
\begin{align*}
 D_1+D_2=\frac{N\lambda_{SH}}{32\pi^2}\left[\sum_{j=h,H^\pm,\chi}m^2_SI_T^1[m^2_j]+v^2\lambda_{SH}I_T^1[m^2_h]+\sum_{j=h,H^\pm,\chi}m^2_jI_T^1[m^2_S]\right].
\end{align*}
The finite, temperature-dependent part of diagram $D_3$ is given in terms of the formulae of eq.~\eqref{eq:ints} as
\begin{align*}
 D_3=&\frac{N \lambda_{SH}}{2}\sum_{j=h,H^\pm,\chi}\left\{I^{-1}_0[m^2_S]I_T^1[m^2_j]+I^{-1}_0[m^2_j]I_T^1[m^2_S]+I_T^0[m^2_S]I_T^0[m^2_j]+I_0^0[m^2_S]I_T^0[m^2_j]\right.\\
 %%%%%
 &\left.+I_0^0[m^2_j]I_T^0[m^2_S]+\left(2\gamma_E+\log\frac{\mu^2}{(4\pi)^2}\right)(I^{-1}_0[m^2_S]I_T^0[m^2_j]+I^{-1}_0[m^2_j]I_T^0[m^2_S])\right\}.
\end{align*}
Finally, the temperature-dependent contribution of the sunset diagram $D_4$ is given by
\begin{align*}
 D_4=-\frac{N v^2\lambda^2_{SH}}{2}(H^0_{T,1}[m^2_S,m^2_h]+H^0_{T,2}[m^2_S,m^2_h]).
\end{align*}
The integral expressions in eq.~\eqref{eq:ints} were obtained by assuming positive masses, and convergence of the integral becomes problematic for negative values. In the scenarios studied, the field-dependent masses for the physical Higgs boson as well as the Goldstone modes become negative for certain values of $h$. To solve this problem, positive quantum corrections to the scalar masses in the Higgs multiplet were resummed by substituting their tree-level values with the one-loop, finite-temperature ones, given in eq.~\eqref{eq:mhresummed}.

\newpage
\bibliographystyle{h-physrev}
\bibliography{bib-EW_barrier}

\end{document}